\pdfoutput=1
\documentclass[acmsmall,screen,nonacm]{acmart}

\renewcommand\footnotetextcopyrightpermission[1]{}
\usepackage[T1]{fontenc}
\usepackage[utf8]{inputenc}
\usepackage{amsmath}
\usepackage{booktabs,longtable,array,calc}
\usepackage{graphicx}
\usepackage{xcolor}
\usepackage{xurl}
\usepackage{float}
\usepackage{microtype}
\usepackage{enumitem}

\providecommand{\tightlist}{\setlength{\itemsep}{0pt}\setlength{\parskip}{0pt}}

\makeatletter
\def\maxwidth{\ifdim\Gin@nat@width>\linewidth\linewidth\else\Gin@nat@width\fi}
\def\maxheight{\ifdim\Gin@nat@height>0.86\textheight 0.86\textheight\else\Gin@nat@height\fi}
\def\fps@figure{htbp}
\makeatother
\setkeys{Gin}{width=\maxwidth,height=\maxheight,keepaspectratio}

\hypersetup{
  pdftitle={From Citation Selection to Citation Absorption: A Measurement Framework for Generative Engine Optimization Across AI Search Platforms},
  pdfauthor={Zhang Kai; He Xinyue; Yao Jingang},
  pdfkeywords={Generative Engine Optimization, AI search, information retrieval, citation influence, retrieval-augmented generation, source attribution, semantic alignment},
  hidelinks
}

\title[From Citation Selection to Citation Absorption]{From Citation Selection to Citation Absorption: A Measurement Framework for Generative Engine Optimization Across AI Search Platforms}
\subtitle{An empirical study of 602 prompts, 21,143 valid search-layer citations, and 23,745 citation-level feature records}

\author{Zhang Kai}
\authornote{First author.}
\affiliation{%
  \institution{Independent Researcher}
  \country{China}
}

\author{He Xinyue}
\authornote{Second author.}
\affiliation{%
  \institution{Independent Researcher}
  \country{China}
}

\author{Yao Jingang}
\authornote{Third author. Public homepage: \url{https://github.com/yaojingang}; X profile: \url{https://x.com/yaojingang}.}
\affiliation{%
  \institution{Independent Researcher}
  \country{China}
}

\date{April 29, 2026}


\begin{abstract}
Generative search engines increasingly determine whether online information is merely discoverable, cited as a source, or actually absorbed into generated answers. This paper proposes and evaluates a two-stage measurement framework for Generative Engine Optimization (GEO): \textbf{citation selection}, where a platform triggers search and chooses sources, and \textbf{citation absorption}, where a cited page contributes language, evidence, structure, or factual support to the final answer. We analyze the public geo-citation-lab dataset, which documents 602 controlled prompts across ChatGPT, Google AI Overview/Gemini, and Perplexity; 21,143 valid search-layer citations; 23,745 citation-level feature records; 18,151 successfully fetched pages; and 72 extracted features. The main descriptive result is a sharp divergence between citation breadth and citation depth. Perplexity cites the most sources on average per prompt, Google also cites broadly, while ChatGPT cites fewer sources but shows substantially higher average citation influence among fetched pages. High-influence pages are longer, more modular, more semantically aligned with the generated answer, and more likely to contain extractable evidence genres such as definitions, numerical facts, comparisons, and procedural steps. A particularly important negative finding is that Q\&A formatting alone does not improve absorption. The empirical pattern supports a view of GEO as \textbf{evidence-container design}: a page must first be eligible for source selection through authority, recognizability, language, and domain context, then useful for absorption through semantic alignment, structural legibility, and evidence density. To preserve scientific validity, this manuscript separates documented descriptive statistics from proposed confirmatory models, avoids causal claims that cannot be supported by the current static snapshot, and includes a claim-level self-audit. The paper contributes a vocabulary and measurement scaffold for studying AI-mediated visibility beyond classical SEO rankings and beyond simple citation counting.
\end{abstract}

\begin{document}
\maketitle

\noindent\textbf{Author and data note.} Author order: Zhang Kai first, He Xinyue second, and Yao Jingang third. Yao Jingang public homepage information: GitHub \url{https://github.com/yaojingang}; X \url{https://x.com/yaojingang}. The public project repository used as the empirical source is \url{https://github.com/yaojingang/geo-citation-lab}. The repository and public report are cited for source traceability.

\section{Counter-Intuitive Findings Placed Before the Main Argument}\label{counter-intuitive-findings-placed-before-the-main-argument}

The dataset produces several results that challenge ordinary GEO intuition. These are placed at the beginning because they determine the correct scientific framing of the paper.

\begin{longtable}[]{@{}
  >{\raggedright\arraybackslash}p{(\columnwidth - 4\tabcolsep) * \real{0.3000}}
  >{\raggedleft\arraybackslash}p{(\columnwidth - 4\tabcolsep) * \real{0.4000}}
  >{\raggedright\arraybackslash}p{(\columnwidth - 4\tabcolsep) * \real{0.3000}}@{}}
\toprule\noalign{}
\begin{minipage}[b]{\linewidth}\raggedright
Finding
\end{minipage} & \begin{minipage}[b]{\linewidth}\raggedleft
Observed evidence
\end{minipage} & \begin{minipage}[b]{\linewidth}\raggedright
Interpretation discipline
\end{minipage} \\
\midrule\noalign{}
\endhead
\bottomrule\noalign{}
\endlastfoot
Citation breadth and citation depth diverge. & Mean citations per prompt: ChatGPT 6.88, Google 12.06, Perplexity 16.35. Mean fetched-page influence: ChatGPT 0.2713, Google 0.0584, Perplexity 0.0646. & Optimize and measure exposure separately from answer-level contribution. \\
Q\&A formatting alone is weak. & Q\&A pages show mean influence 0.0947 versus 0.1005 for non-Q\&A pages, a -5.74\% relative difference. & Question-answer packaging is insufficient without evidence density and semantic fit. \\
News is selected often, yet news\_media is not the deepest absorbed domain type. & News appears frequently in source selection, while news\_media pages average 0.0726 influence and encyclopedia pages average 0.2144 in the reported domain-type table. & Selection probability and absorption intensity are separate outcomes. \\
English does not universally increase citation breadth. & In the C-layer language contrast, ChatGPT has higher average citation count for Chinese prompts, 7.77 versus 7.03 for English. Google shows the opposite, 11.57 English versus 7.53 Chinese. & Language effects should be modeled with platform interactions. \\
More constraints can reduce citation breadth on one platform. & For D-layer multi-constraint tasks, ChatGPT averages 3.4 citations while Google averages 12.6 and Perplexity 17.7. & Complexity can trigger compression and internal synthesis rather than broader retrieval. \\
\end{longtable}

\textbf{Evidence standard.} This manuscript avoids fabricated p-values, confidence intervals, or regression coefficients. All numerical claims in the results section are descriptive and traceable to the public geo-citation-lab report and repository. Confirmatory inference specifications are reported as a transparent analysis plan for future raw-CSV reruns under an explicit analysis script.

\section{Introduction}\label{introduction}

Search visibility used to be modeled primarily as a ranked-list problem. A user issued a query, a search engine returned links, and the publisher sought high placement, click-through, and conversion. Generative engines modify every step of this loop. They retrieve or consult sources, synthesize an answer, attach citations, and often satisfy the user before any source is opened. This changes the unit of measurement from ranking position to answer participation.

The practical question for GEO is therefore not limited to whether a source appears in a citation list. A page can be cited as a weak navigational reference while another page supplies definitions, numerical evidence, comparisons, or procedural steps that shape multiple paragraphs of the generated answer. This paper names the first outcome \textbf{citation selection} and the second outcome \textbf{citation absorption}.

The distinction is theoretically important because the same observed citation can play different epistemic roles. It can serve as background, a factual basis, a paraphrase source, a structural guide, a direct quote, or a low-impact reference. It is also practically important because publishers and researchers need to know whether optimization should target authority and indexability, content structure and semantic alignment, or both.

The public geo-citation-lab dataset creates an unusually useful research object for this question. It contains controlled prompts, multiple generative search platforms, cleaned search-layer citation records, fetched-page status, and citation-level features. Its design allows the study of search triggering, source selection, and source absorption under one pipeline.

This paper is framed as an empirical measurement study. It does not claim access to hidden model internals, and it does not claim that observed content features causally force a generative engine to cite or use a page. It establishes a descriptive framework, documents reproducible patterns, and specifies the inferential models required for a causal or quasi-causal extension.

\subsection{Contributions}\label{contributions}

\begin{enumerate}
\def\labelenumi{\arabic{enumi}.}
\tightlist
\item
  A two-stage formalization of GEO that separates citation selection from citation absorption.
\item
  A cross-platform empirical summary of ChatGPT, Google AI Overview/Gemini, and Perplexity using the public geo-citation-lab dataset.
\item
  A measurement interpretation of \texttt{influence\_score} as an answer-level absorption proxy, including its mathematical components and the modeling restrictions that follow from its construction.
\item
  A set of counter-intuitive empirical findings that challenge shallow GEO heuristics such as maximizing citation count or converting all content into Q\&A pages.
\item
  A scientific self-audit and reproducibility checklist designed to support independent review and replication.
\end{enumerate}

\section{Related Work}\label{related-work}

\subsection{Generative Engine Optimization}\label{generative-engine-optimization}

GEO was formalized by Aggarwal et al.~as a framework for improving content visibility in generative engine responses {[}1{]}. That work introduced GEO-bench and showed that black-box content interventions could improve visibility under specific benchmark conditions. The present paper differs in its measurement target: it studies observed cross-platform citation behavior and differentiates a source being selected from a source being deeply used.

Recent work has extended GEO toward AI search benchmarking, document-centric optimization, and diagnosis of citation failures {[}2,3,4,5,6{]}. Citation-repair work emphasizes that failures can occur across retrieval, fetching, parsing, attribution, and generation {[}3{]}. Agentic and structural GEO work further supports the premise that citation outcomes should be studied as a pipeline rather than as a single ranking event {[}5,6{]}.

\subsection{Generative Search, RAG, and Source Attribution}\label{generative-search-rag-and-source-attribution}

Generative search inherits elements from retrieval-augmented generation: retrieve candidate material, ground a response, and provide attribution {[}12,13,14,15{]}. Deployed answer engines introduce additional uncertainty because source selection, answer generation, and citation rendering are often black-box platform decisions. Studies of answer engines and source-cited responses document limitations such as hallucination, inaccurate citation, and mismatch between evidence and generated claims {[}7,8,9{]}. Citation-evaluation papers further show that citation quality needs sentence-level or claim-level analysis, since a source link can be present while failing to support the statement it is attached to {[}9{]}.

This paper is complementary to attribution-quality research. It does not evaluate whether every generated claim is fully supported. Its unit of analysis is the source page: how often it is selected, how it is characterized, and how much it appears to influence the answer text and structure.

\subsection{Why GEO Needs a New Dependent Variable}\label{why-geo-needs-a-new-dependent-variable}

Classical SEO metrics emphasize ranking, click-through, backlinks, domain strength, and user engagement. In generative search, those variables can still influence whether a source enters the retrieval pool, yet they do not directly measure whether the answer depends on that source. A generative engine can cite a high-authority page as a reference while extracting substantive facts from a lower-frequency but more structured page.

The conceptual move in this paper is to define answer-level source influence as a second dependent variable. It is imperfect, because the engine's internal attention and retrieval traces are unavailable, but it can be approximated through observable properties such as repeated references, position, paragraph coverage, text overlap, and semantic alignment.

\section{Dataset and Provenance}\label{dataset-and-provenance}

\subsection{Project Scope}\label{project-scope}

The geo-citation-lab repository describes itself as a dataset and analysis pipeline for studying how AI search engines select and use citations {[}16{]}. The repository README identifies the original research author as Zhang Kai and the secondary report/open-source organization role as Yao Jingang {[}16{]}. This manuscript lists Zhang Kai as first author, He Xinyue as second author, and Yao Jingang as third author. Yao Jingang's public GitHub and X pages are included in the author note and references {[}18,19{]}. Repository-maintainer attribution is retained in the data-source references.

The public snapshot includes 602 prompts, three platforms, 21,181 cleaned search-layer rows, 21,143 valid search-layer citations, 23,745 citation-level feature rows, 72 feature dimensions, and 18,151 successfully fetched citation pages. The reported overall fetch success rate is 76.44\% {[}16,17{]}.

\begin{longtable}[]{@{}
  >{\raggedright\arraybackslash}p{(\columnwidth - 4\tabcolsep) * \real{0.3000}}
  >{\raggedleft\arraybackslash}p{(\columnwidth - 4\tabcolsep) * \real{0.4000}}
  >{\raggedright\arraybackslash}p{(\columnwidth - 4\tabcolsep) * \real{0.3000}}@{}}
\toprule\noalign{}
\begin{minipage}[b]{\linewidth}\raggedright
Asset
\end{minipage} & \begin{minipage}[b]{\linewidth}\raggedleft
Reported quantity
\end{minipage} & \begin{minipage}[b]{\linewidth}\raggedright
Analytical role
\end{minipage} \\
\midrule\noalign{}
\endhead
\bottomrule\noalign{}
\endlastfoot
Prompts & 602 total & Controlled inputs for search triggering and source selection. \\
Prompt layers & A/B/C/D = 432/60/60/50 & Main experiment, style contrast, language contrast, and realistic/extreme scenarios. \\
Platforms & 3 & ChatGPT, Google AI Overview/Gemini, and Perplexity. \\
Cleaned search-layer rows & 21,181 & Search-trigger and citation-domain evidence. \\
Valid search-layer citations & 21,143 & Primary source-selection sample. \\
Citation-level feature rows & 23,745 & Feature table used for influence analysis. \\
Fetch-ok pages & 18,151 & Primary absorption sample. \\
Feature dimensions & 72 & Page structure, genre, semantic alignment, LLM ratings, and outcome features. \\
\end{longtable}

\subsection{Prompt Architecture}\label{prompt-architecture}

\begin{longtable}[]{@{}
  >{\raggedright\arraybackslash}p{(\columnwidth - 6\tabcolsep) * \real{0.2308}}
  >{\raggedleft\arraybackslash}p{(\columnwidth - 6\tabcolsep) * \real{0.3077}}
  >{\raggedright\arraybackslash}p{(\columnwidth - 6\tabcolsep) * \real{0.2308}}
  >{\raggedright\arraybackslash}p{(\columnwidth - 6\tabcolsep) * \real{0.2308}}@{}}
\toprule\noalign{}
\begin{minipage}[b]{\linewidth}\raggedright
Layer
\end{minipage} & \begin{minipage}[b]{\linewidth}\raggedleft
N
\end{minipage} & \begin{minipage}[b]{\linewidth}\raggedright
Purpose
\end{minipage} & \begin{minipage}[b]{\linewidth}\raggedright
Controlled or observed factors
\end{minipage} \\
\midrule\noalign{}
\endhead
\bottomrule\noalign{}
\endlastfoot
A & 432 & Main experimental layer & Task type, trigger strength, time sensitivity, industry, subtask. \\
B & 60 & Style contrast & Natural phrasing, explicit source request, expert-role prompt. \\
C & 60 & Language contrast & Chinese-English pair design. \\
D & 50 & Realistic and extreme scenarios & High-risk, ambiguous, multi-constraint, long-decision, macro-trend cases. \\
\end{longtable}

This design is stronger than an uncontrolled scrape because it creates repeated conditions under which prompt type, phrasing, language, and scenario complexity can be compared. At the same time, the sample is not a probability sample of all real user traffic. All external validity claims must therefore be bounded to the prompt distribution used here.

\subsection{Cleaning and Caveats}\label{cleaning-and-caveats}

The public repository and report document several cleaning caveats: repeated header rows in the ChatGPT CSV, normalization issues for ChatGPT A\_news and A\_technology naming, 15 missing ChatGPT prompt outputs after cleaning, noisy website-type values, and unknown values in country and language fields {[}16,17{]}. These details matter because cross-platform comparisons can be distorted if denominators are silently changed.

The analysis in this manuscript uses the report's cleaned denominators: 587 observed ChatGPT prompts, 602 Google prompts, and 602 Perplexity prompts for search-layer platform summaries. Citation absorption is restricted to \texttt{fetch\_ok} pages where page content was successfully retrieved.

\section{Measurement Framework}\label{measurement-framework}

\begin{figure}
\centering
\includegraphics[width=\linewidth]{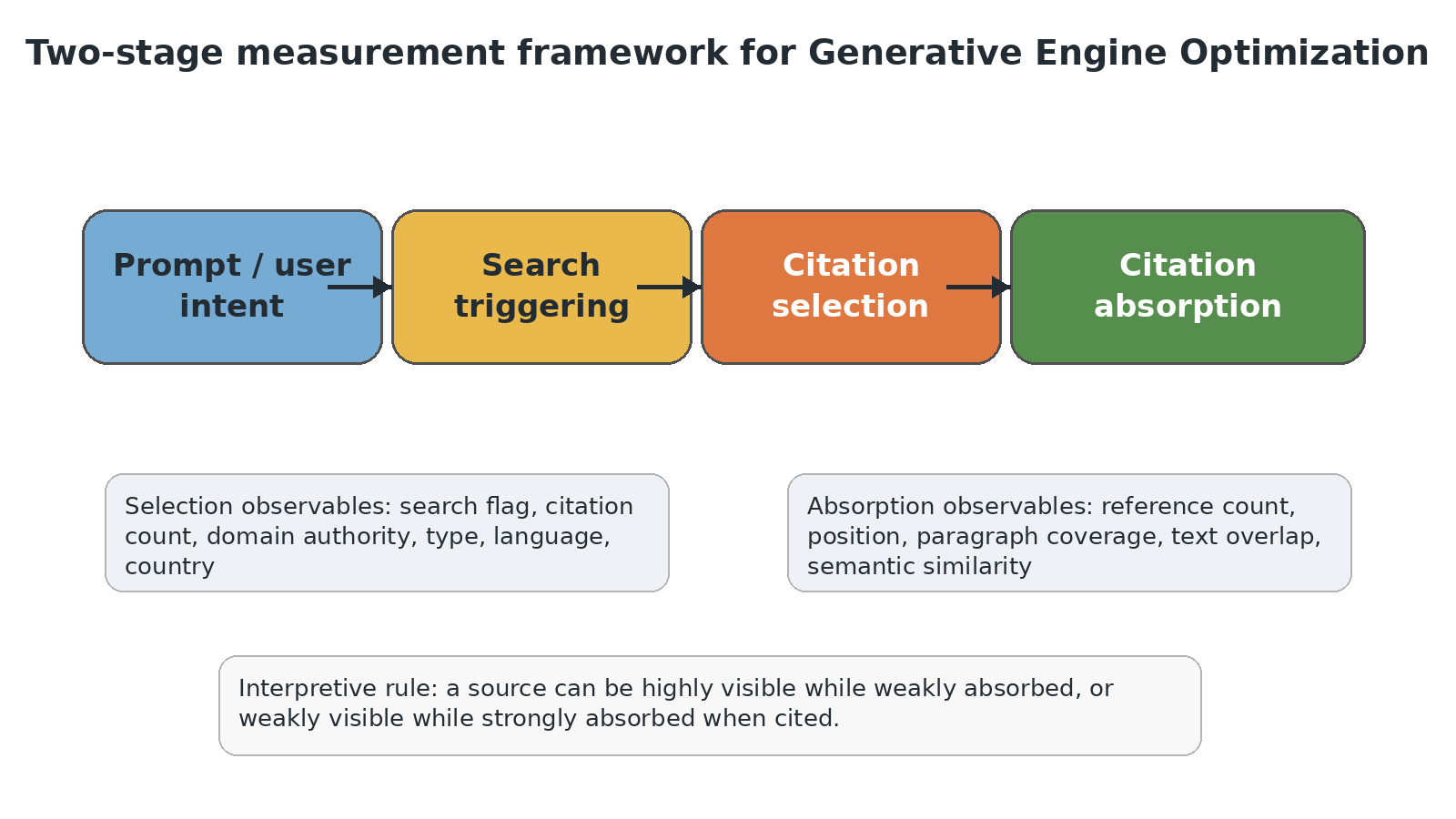}
\caption{The GEO measurement problem decomposes into citation selection and citation absorption. Source: author synthesis from the geo-citation-lab pipeline design.}
\end{figure}

\subsection{Citation Selection}\label{citation-selection}

For prompt \(p\) and platform \(e\), define a search-trigger variable \(S_{p,e}\) and a citation count \(C_{p,e}\). The public report computes prompt-level triggering by treating a prompt as triggered if any cleaned row under the prompt indicates search was triggered. Citation count is the count of valid citation-domain rows under the same prompt-platform observation.

\[
S_{p,e}=\mathbb{1}\{\text{any cleaned row for prompt }p\text{ on platform }e\text{ indicates search triggered}\}
\]

\[
C_{p,e}=\sum_i ValidCitation_{i,p,e}
\]

Equation (1) defines selection outcomes. The citation count is a breadth measure and should not be interpreted as an absorption measure.

\subsection{Citation Absorption}\label{citation-absorption}

Citation absorption is measured at the citation-page level. The public feature table computes an \texttt{influence\_score} as a weighted proxy for how deeply a cited page appears to shape a generated answer. The score rewards repeated reference, early appearance, coverage across answer paragraphs, TF-IDF similarity, and n-gram overlap {[}17{]}.

\[
\begin{aligned}
Influence_i = &\;0.20\cdot \min(ref\_count_i/3,1) \\
&+0.15\cdot (1-first\_position\_ratio_i) \\
&+0.20\cdot paragraph\_coverage\_ratio_i \\
&+0.25\cdot tfidf\_cosine_i \\
&+0.20\cdot \frac{bigram\_overlap_i+trigram\_overlap_i}{2}.
\end{aligned}
\]

Equation (2) is the absorption proxy used in the public report. Since these components define the outcome, they must not be reused as independent causal predictors of the same score.

\subsection{Scientific Interpretation of the Influence Score}\label{scientific-interpretation-of-the-influence-score}

The influence score is a constructed observational proxy rather than a direct measure of hidden model attention, retrieval ranking, or causal dependence. It is useful because it connects source-page content to answer-level textual evidence, but it must be interpreted with discipline.

The most important modeling rule is outcome-component separation. Variables used inside Equation (2), such as \texttt{ref\_count}, \texttt{first\_position\_ratio}, \texttt{paragraph\_coverage\_ratio}, TF-IDF similarity, and n-gram overlap, can be described as parts of influence. They should not be claimed as independent drivers of influence in a regression that uses \texttt{influence\_score} as the dependent variable.

Cleaner explanatory variables include page structure, evidence genre, title-question match, embedding similarity measures that are not directly used in the score, LLM relevance score, LLM quality score, domain type, source type, language, and platform interactions. Even these variables remain observational unless manipulated experimentally.

\subsection{Estimands and Model Specifications for a Raw CSV Rerun}\label{estimands-and-model-specifications-for-a-raw-csv-rerun}

This manuscript reports descriptive estimands and specifies confirmatory models for an expanded microdata analysis. The confirmatory models should be populated only after re-running the raw CSVs with a locked script, exporting model diagnostics, and documenting missingness.

For selection breadth:

\[
C_{p,e}\sim NegativeBinomial(\mu_{p,e},\alpha)
\]

\[
\log\mu_{p,e}=\beta_0+\beta_{platform[e]}+\beta_{layer[p]}+\beta_{style[p]}+\beta_{language[p]}+\beta_{industry[p]}+\beta_{interactions}+u_p.
\]

For absorption:

\[
\operatorname{logit}(\mathbb{E}[Influence_i])=\gamma_0+\gamma_{platform[e_i]}+\gamma_{domain\_type[d_i]}+\gamma_{structure}^T X_i+\gamma_{genre}^T G_i+\gamma_{semantic}^T Z_i+\epsilon_i.
\]

For the absorption model, fractional logit or beta regression can be used, with two-way clustered standard errors by prompt and domain. Variables that are direct components of Equation (2) should be excluded from the primary explanatory specification.

\subsection{Identification Map: What Can and Cannot Be Claimed}\label{identification-map-what-can-and-cannot-be-claimed}

A rigorous GEO paper should make the identification status of every claim explicit. The present dataset supports strong descriptive claims about observed platform behavior under the prompt design. It supports structured hypotheses about mechanisms, because multiple independent descriptive patterns point in the same direction. It does not, by itself, support universal causal claims about how changing one page feature will necessarily change citation outcomes on a live platform.

The identification map has four levels. Level 1 claims are direct counts: number of prompts, number of citations, trigger rates, and platform-level means. Level 2 claims are descriptive contrasts: differences in citation breadth across platforms, differences in influence by domain type, and top-versus-bottom quartile comparisons. Level 3 claims are mechanistic interpretations: evidence-container design, platform compression, and selection-absorption separation. Level 4 claims are causal optimization prescriptions: for example, a statement that adding definitions will increase future AI citations for a specific page. This paper only treats Levels 1 and 2 as empirical findings. Level 3 is presented as an explanation consistent with the data. Level 4 is reserved for future intervention experiments.

This hierarchy is essential because GEO research is likely to be commercially sensitive. It is tempting to translate every correlation into a content tactic. That translation can be useful as a hypothesis-generation tool, yet it should not be marketed as a scientific law unless the feature has been manipulated under controlled conditions. The dataset strongly suggests that semantic alignment and extractable evidence matter, but it cannot isolate whether headings cause absorption, whether high-quality pages simply tend to have more headings, or whether both are downstream of a latent editorial-quality variable.

\begin{longtable}[]{@{}
  >{\raggedright\arraybackslash}p{(\columnwidth - 6\tabcolsep) * \real{0.2500}}
  >{\raggedright\arraybackslash}p{(\columnwidth - 6\tabcolsep) * \real{0.2500}}
  >{\raggedright\arraybackslash}p{(\columnwidth - 6\tabcolsep) * \real{0.2500}}
  >{\raggedright\arraybackslash}p{(\columnwidth - 6\tabcolsep) * \real{0.2500}}@{}}
\toprule\noalign{}
\begin{minipage}[b]{\linewidth}\raggedright
Claim level
\end{minipage} & \begin{minipage}[b]{\linewidth}\raggedright
Example
\end{minipage} & \begin{minipage}[b]{\linewidth}\raggedright
Supported by this manuscript?
\end{minipage} & \begin{minipage}[b]{\linewidth}\raggedright
Recommended evidence for stronger claim
\end{minipage} \\
\midrule\noalign{}
\endhead
\bottomrule\noalign{}
\endlastfoot
Level 1: direct count & Perplexity has 602 triggered prompts in the cleaned report. & Yes. & Exact CSV reproduction and timestamp logging. \\
Level 2: descriptive contrast & ChatGPT has lower citation breadth and higher fetched-page mean influence. & Yes. & Bootstrap confidence intervals and denominator audit. \\
Level 3: mechanism & High-influence pages function as evidence containers. & Plausible interpretation. & Multivariate controls and qualitative page inspection. \\
Level 4: causal optimization & Adding comparison sections will increase future absorption. & Not established here. & Randomized page rewrites and repeated platform querying. \\
\end{longtable}

\section{Results: Search Triggering and Citation Breadth}\label{results-search-triggering-and-citation-breadth}

\begin{figure}
\centering
\includegraphics[width=\linewidth]{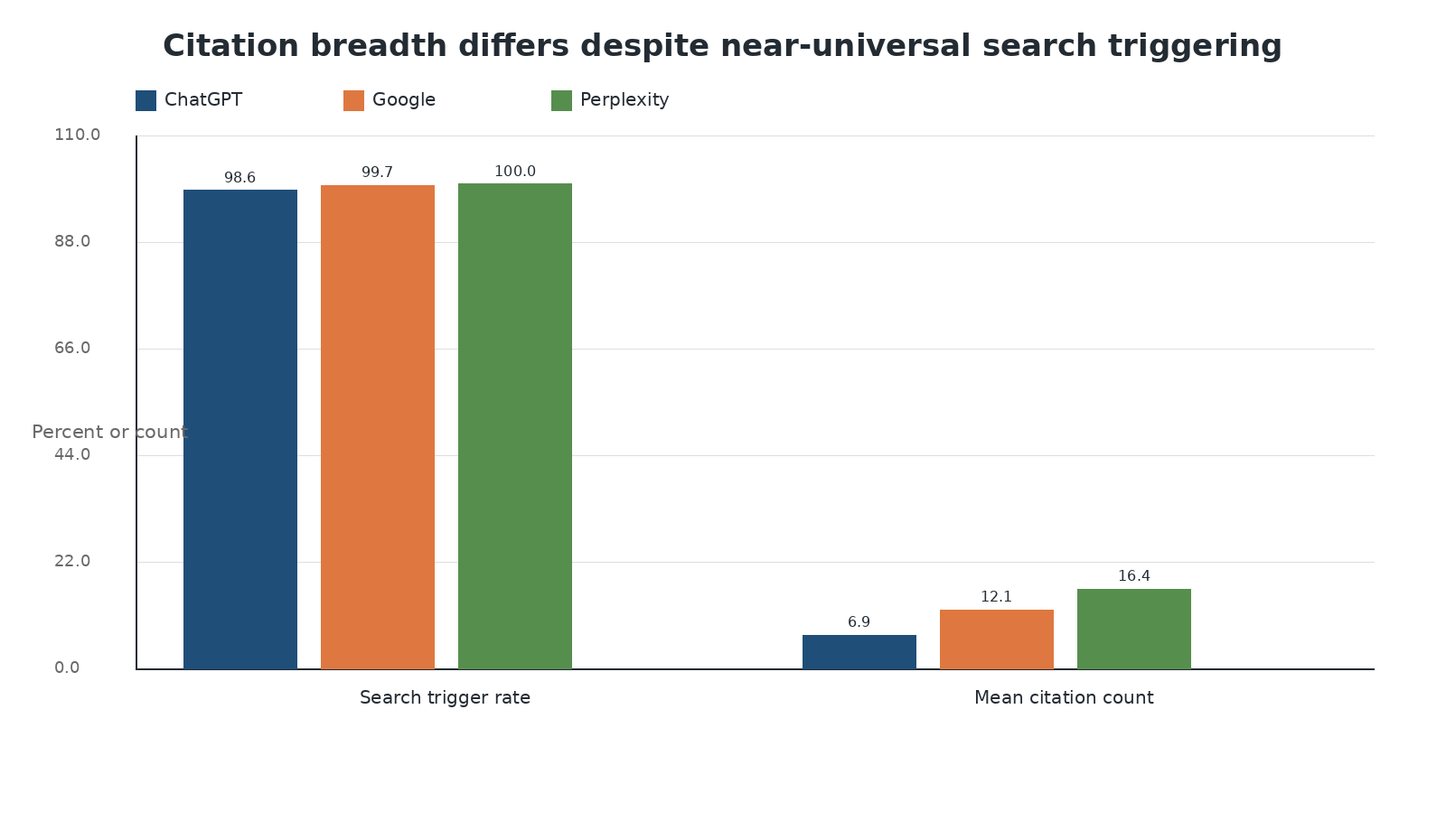}
\caption{Search triggering is near universal, but citation breadth differs sharply by platform. Source: geo-citation-lab public report.}
\end{figure}

\begin{longtable}[]{@{}
  >{\raggedright\arraybackslash}p{(\columnwidth - 12\tabcolsep) * \real{0.1111}}
  >{\raggedleft\arraybackslash}p{(\columnwidth - 12\tabcolsep) * \real{0.1481}}
  >{\raggedleft\arraybackslash}p{(\columnwidth - 12\tabcolsep) * \real{0.1481}}
  >{\raggedleft\arraybackslash}p{(\columnwidth - 12\tabcolsep) * \real{0.1481}}
  >{\raggedleft\arraybackslash}p{(\columnwidth - 12\tabcolsep) * \real{0.1481}}
  >{\raggedleft\arraybackslash}p{(\columnwidth - 12\tabcolsep) * \real{0.1481}}
  >{\raggedleft\arraybackslash}p{(\columnwidth - 12\tabcolsep) * \real{0.1481}}@{}}
\toprule\noalign{}
\begin{minipage}[b]{\linewidth}\raggedright
Platform
\end{minipage} & \begin{minipage}[b]{\linewidth}\raggedleft
Observed prompts
\end{minipage} & \begin{minipage}[b]{\linewidth}\raggedleft
Triggered prompts
\end{minipage} & \begin{minipage}[b]{\linewidth}\raggedleft
Trigger rate
\end{minipage} & \begin{minipage}[b]{\linewidth}\raggedleft
Mean citations
\end{minipage} & \begin{minipage}[b]{\linewidth}\raggedleft
Median citations
\end{minipage} & \begin{minipage}[b]{\linewidth}\raggedleft
Max citations
\end{minipage} \\
\midrule\noalign{}
\endhead
\bottomrule\noalign{}
\endlastfoot
ChatGPT & 587 & 579 & 98.64\% & 6.88 & 6 & 21 \\
Google AIO & 602 & 600 & 99.67\% & 12.06 & 12 & 37 \\
Perplexity & 602 & 602 & 100.00\% & 16.35 & 17 & 27 \\
\end{longtable}

The near-ceiling search-trigger rates imply that the primary frontier in this dataset is no longer whether generative engines search. The main differences lie in how many sources are selected and how those sources are used. Perplexity is the broadest citing platform, Google is also broad, and ChatGPT is citation-sparser.

A naive GEO strategy might treat broader citation as better. That conclusion would be premature because the absorption results below show that fewer citations can coexist with deeper use of each citation. The correct unit of analysis depends on the objective: traffic attribution, source exposure, claim support, or answer influence.

\subsection{Prompt Style and Language Effects}\label{prompt-style-and-language-effects}

\begin{longtable}[]{@{}lrrr@{}}
\toprule\noalign{}
Platform & Natural & Explicit source request & Expert-role prompt \\
\midrule\noalign{}
\endhead
\bottomrule\noalign{}
\endlastfoot
ChatGPT & 7.30 & 6.15 & 7.95 \\
Google & 14.05 & 15.90 & 10.40 \\
Perplexity & 15.70 & 17.15 & 16.70 \\
\end{longtable}

\begin{longtable}[]{@{}lrrl@{}}
\toprule\noalign{}
Platform & Chinese prompt & English prompt & Observed pattern \\
\midrule\noalign{}
\endhead
\bottomrule\noalign{}
\endlastfoot
ChatGPT & 7.77 & 7.03 & Chinese higher in this sample. \\
Google & 7.53 & 11.57 & English substantially higher. \\
Perplexity & 15.93 & 16.43 & English slightly higher. \\
\end{longtable}

Prompt style effects are platform-specific. Google and Perplexity respond strongly to explicit source requests, while ChatGPT shows higher average citation count under expert-role prompts in this layer. This pattern argues against a universal prompt-engineering rule for GEO measurement.

The language contrast is equally important. English prompts increase citation breadth on Google and slightly on Perplexity, but ChatGPT shows higher average citation count for Chinese prompts in the reported C-layer sample. Language should therefore be modeled with platform interactions rather than as a global treatment.

\subsection{Realistic and Extreme Scenarios}\label{realistic-and-extreme-scenarios}

\begin{longtable}[]{@{}lrrr@{}}
\toprule\noalign{}
Scenario & ChatGPT & Google & Perplexity \\
\midrule\noalign{}
\endhead
\bottomrule\noalign{}
\endlastfoot
High-risk & 6.0 & 13.9 & 16.0 \\
Ambiguous & 7.9 & 8.9 & 13.1 \\
Multi-constraint & 3.4 & 12.6 & 17.7 \\
Long-decision & 9.2 & 14.5 & 17.4 \\
Macro-trend & 8.6 & 13.4 & 15.1 \\
\end{longtable}

The strongest scenario contrast is multi-constraint behavior. ChatGPT cites only 3.4 sources on average, while Perplexity cites 17.7. One interpretation is that ChatGPT compresses complex constraints into internal synthesis after selecting fewer sources, while Perplexity decomposes the task into broader retrieval. This remains an interpretation, not a causal claim, until the prompt-level outputs are inspected and modeled.

\section{Results: Source Selection}\label{results-source-selection}

\begin{figure}
\centering
\includegraphics[width=\linewidth]{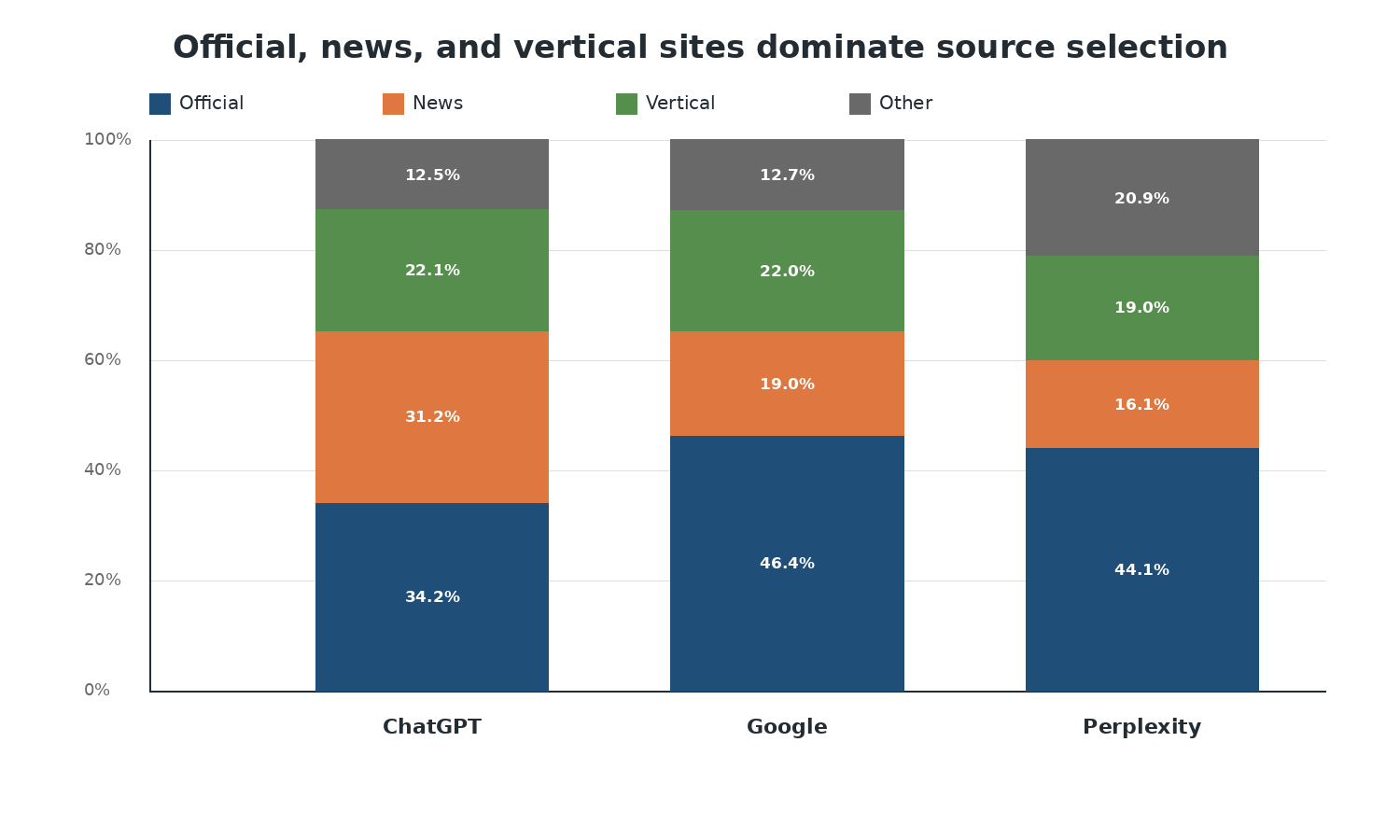}
\caption{Source-type composition indicates that official, news, and vertical sources form the default candidate pool. Source: geo-citation-lab public report.}
\end{figure}

\begin{longtable}[]{@{}lrrrr@{}}
\toprule\noalign{}
Platform & Official & News & Vertical & Official + News + Vertical \\
\midrule\noalign{}
\endhead
\bottomrule\noalign{}
\endlastfoot
ChatGPT & 34.22\% & 31.17\% & 22.13\% & 87.52\% \\
Google & 46.35\% & 18.99\% & 22.00\% & 87.34\% \\
Perplexity & 44.07\% & 16.07\% & 18.99\% & 79.12\% \\
\end{longtable}

The source-selection layer is highly concentrated. Official, news, and vertical sources account for 79.12\% to 87.52\% of citations across platforms. This concentration suggests that source identity remains a strong entry condition. However, identity does not determine absorption intensity, as shown later by the lower average influence for \texttt{news\_media} relative to encyclopedia pages in the reported domain-type table.

\begin{longtable}[]{@{}
  >{\raggedright\arraybackslash}p{(\columnwidth - 8\tabcolsep) * \real{0.1579}}
  >{\raggedleft\arraybackslash}p{(\columnwidth - 8\tabcolsep) * \real{0.2105}}
  >{\raggedleft\arraybackslash}p{(\columnwidth - 8\tabcolsep) * \real{0.2105}}
  >{\raggedleft\arraybackslash}p{(\columnwidth - 8\tabcolsep) * \real{0.2105}}
  >{\raggedleft\arraybackslash}p{(\columnwidth - 8\tabcolsep) * \real{0.2105}}@{}}
\toprule\noalign{}
\begin{minipage}[b]{\linewidth}\raggedright
Platform
\end{minipage} & \begin{minipage}[b]{\linewidth}\raggedleft
US share
\end{minipage} & \begin{minipage}[b]{\linewidth}\raggedleft
English share
\end{minipage} & \begin{minipage}[b]{\linewidth}\raggedleft
Mean Final\_DR
\end{minipage} & \begin{minipage}[b]{\linewidth}\raggedleft
Median Final\_DR
\end{minipage} \\
\midrule\noalign{}
\endhead
\bottomrule\noalign{}
\endlastfoot
ChatGPT & 85.89\% & 95.07\% & 584.60 & 592 \\
Google & 86.76\% & 91.98\% & 541.15 & 526 \\
Perplexity & 82.70\% & 82.90\% & 558.33 & 542 \\
\end{longtable}

The identified country and language samples are dominated by US and English sources. The report warns that country and language fields contain unknown values, so these percentages should be read as identifiable-sample shares. Final\_DR medians between 526 and 592 indicate that authority-like signals remain important for entering the candidate pool.

\begin{longtable}[]{@{}rlrl@{}}
\toprule\noalign{}
Rank & Domain & Count & Interpretive category \\
\midrule\noalign{}
\endhead
\bottomrule\noalign{}
\endlastfoot
1 & youtube.com & 560 & Platform / video aggregation \\
2 & en.wikipedia.org & 352 & Encyclopedic explanation \\
3 & reddit.com & 315 & Community / forum \\
4 & reuters.com & 287 & Wire service \\
5 & linkedin.com & 187 & Professional platform \\
6 & nytimes.com & 174 & News media \\
7 & pmc.ncbi.nlm.nih.gov & 167 & Biomedical literature \\
8 & facebook.com & 151 & Social platform \\
9 & forbes.com & 146 & Business media \\
10 & finance.yahoo.com & 146 & Finance portal \\
11 & deloitte.com & 134 & Professional services \\
12 & theguardian.com & 124 & News media \\
13 & wsj.com & 122 & News media \\
14 & investopedia.com & 121 & Finance explainer \\
15 & weforum.org & 121 & Institutional thought leadership \\
\end{longtable}

High-frequency domains reveal which sources are repeatedly eligible for selection. They do not prove which domains most strongly shape answers. This distinction is central to the selection-absorption separation.

\section{Results: Citation Absorption}\label{results-citation-absorption}

\begin{figure}
\centering
\includegraphics[width=\linewidth]{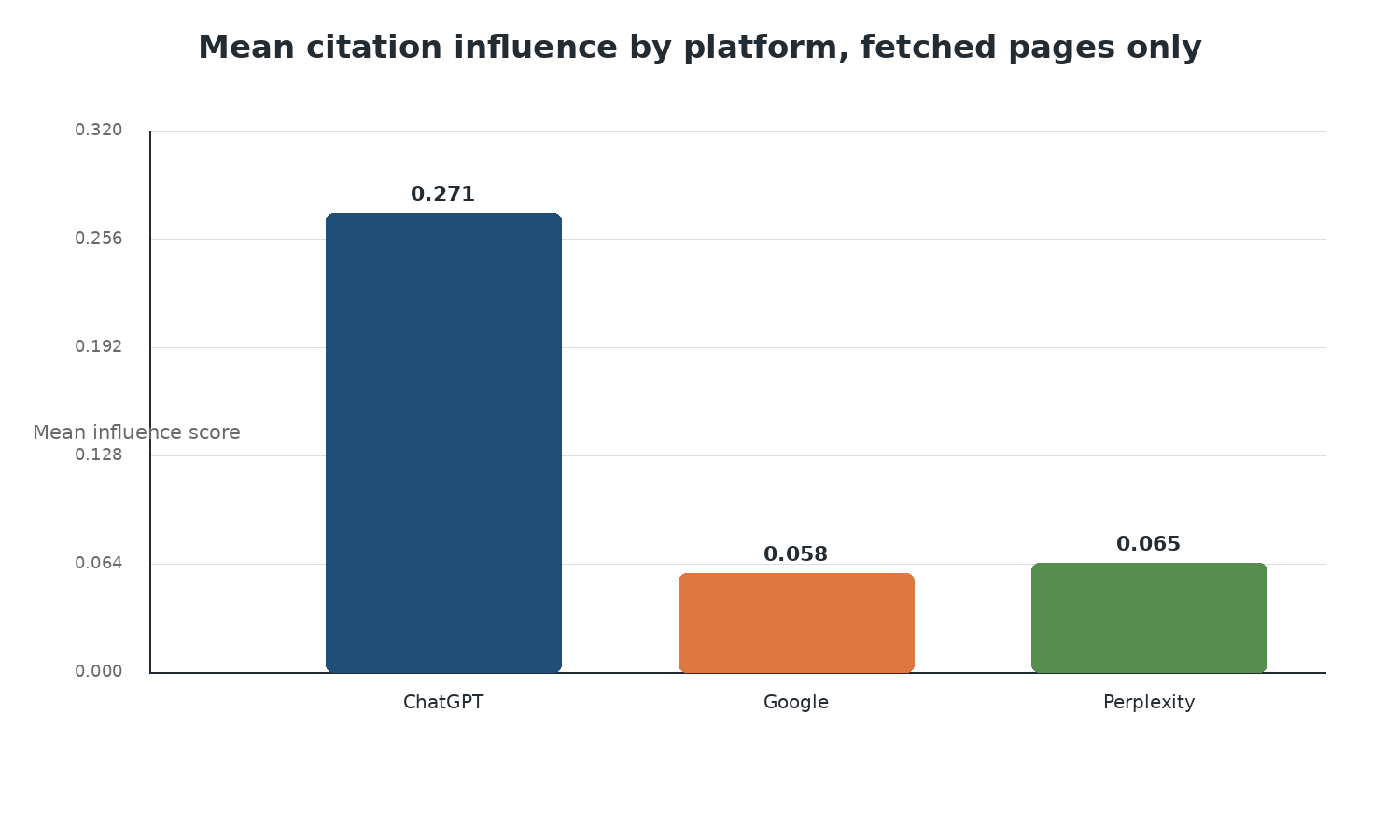}
\caption{ChatGPT has substantially higher mean citation influence among fetched pages. Source: geo-citation-lab public report.}
\end{figure}

\begin{longtable}[]{@{}lrrr@{}}
\toprule\noalign{}
Platform & Fetch-ok citations & Mean influence & Median influence \\
\midrule\noalign{}
\endhead
\bottomrule\noalign{}
\endlastfoot
ChatGPT & 3,323 & 0.2713 & 0.2611 \\
Google & 6,385 & 0.0584 & 0.0515 \\
Perplexity & 8,443 & 0.0646 & 0.0333 \\
\end{longtable}

This result is the core empirical motivation for this paper. ChatGPT cites fewer sources but uses the sources it cites more deeply according to the influence proxy. Perplexity cites the most sources, yet its mean influence is much lower. Google sits between the two in citation breadth but closer to Perplexity in citation absorption.

This finding changes the GEO objective function. For traffic-oriented attribution, a broad citation list may be valuable. For answer-shaping power, a smaller number of high-absorption citations may matter more. Any single metric of visibility collapses these objectives and loses important information.

\subsection{Page Structure and Length}\label{page-structure-and-length}

\begin{figure}
\centering
\includegraphics[width=\linewidth]{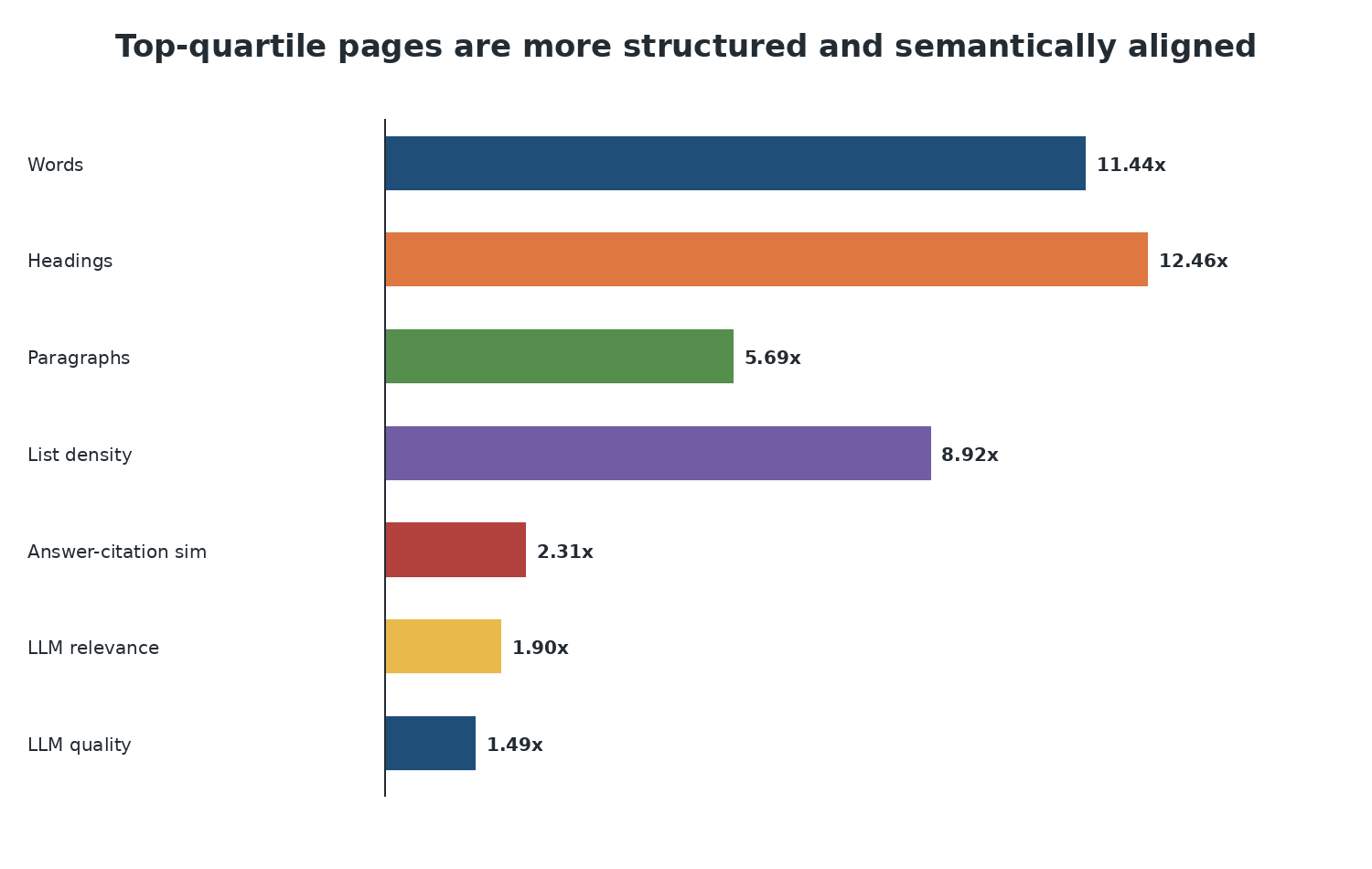}
\caption{Ratio of top-quartile to bottom-quartile page attributes by influence score. Source: geo-citation-lab public report.}
\end{figure}

\begin{longtable}[]{@{}lrrr@{}}
\toprule\noalign{}
Metric & Top 25\% & Bottom 25\% & Ratio \\
\midrule\noalign{}
\endhead
\bottomrule\noalign{}
\endlastfoot
Word count & 1,943.30 & 169.82 & 11.44x \\
Heading total & 10.59 & 0.85 & 12.50x \\
Paragraph count & 47.49 & 8.34 & 5.69x \\
List density & 0.428 & 0.048 & 8.94x \\
Answer-citation semantic similarity & 0.570 & 0.247 & 2.31x \\
LLM relevance score & 3.535 & 1.856 & 1.90x \\
LLM content quality & 3.404 & 2.289 & 1.49x \\
\end{longtable}

High-influence pages are not merely longer. They contain more headings, more paragraphs, denser list structures, higher semantic similarity, and higher LLM-rated relevance and quality. The observed pattern is consistent with the evidence-container hypothesis: pages that package multiple extractable units can be reused across more answer segments.

\begin{figure}
\centering
\includegraphics[width=\linewidth]{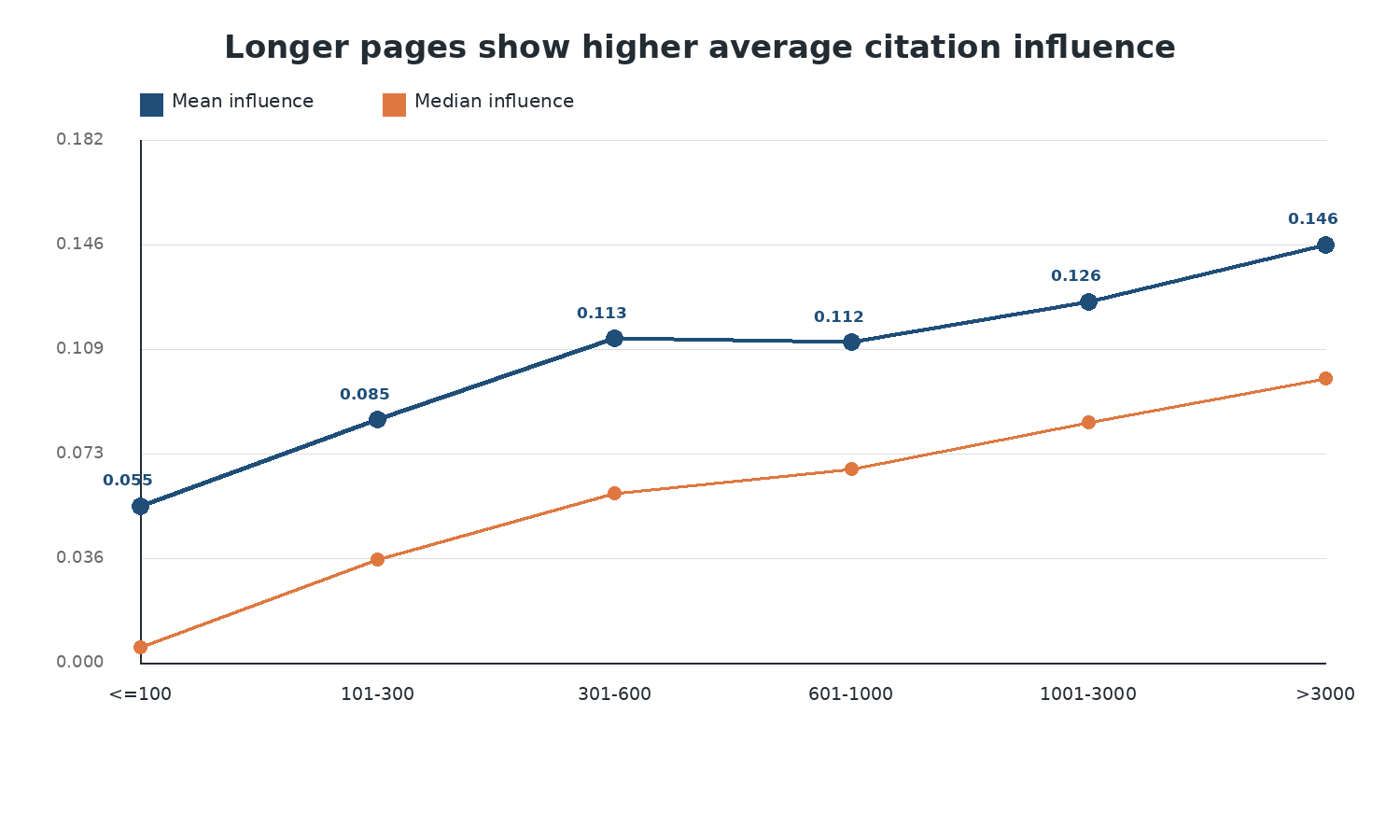}
\caption{Mean and median influence by word-count bin. Source: geo-citation-lab public report.}
\end{figure}

The word-count bin analysis shows monotonic improvement in mean influence from very short pages to pages longer than 3,000 words, with a modest local plateau around 301 to 1,000 words. This should not be reduced to a rule that longer content always wins. Length appears beneficial when it is coupled with usable structure, semantic alignment, and evidence density.

\subsection{Semantic Alignment and Evidence Genres}\label{semantic-alignment-and-evidence-genres}

\begin{figure}
\centering
\includegraphics[width=\linewidth]{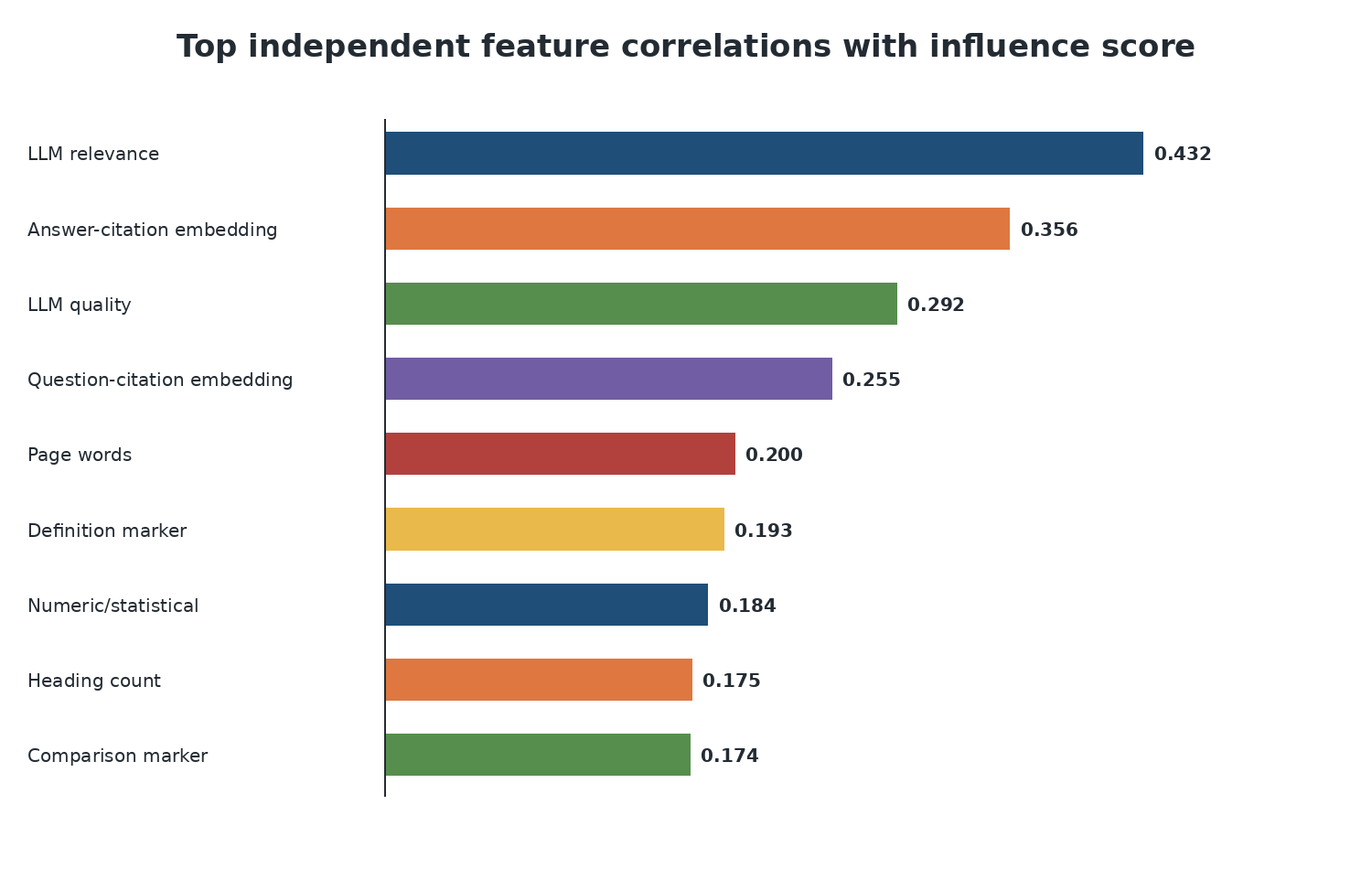}
\caption{Reported correlations between independent features and influence\_score. Variables used directly in the score definition are intentionally omitted from this figure. Source: geo-citation-lab public report.}
\end{figure}

The strongest reported independent correlation is LLM relevance score (\(r=0.4322\)), followed by answer-citation embedding similarity (\(r=0.3561\)), LLM content quality (\(r=0.2917\)), and question-citation embedding similarity (\(r=0.2548\)). Page word count and structural markers matter, but semantic fit is stronger than a simple length signal.

\begin{figure}
\centering
\includegraphics[width=\linewidth]{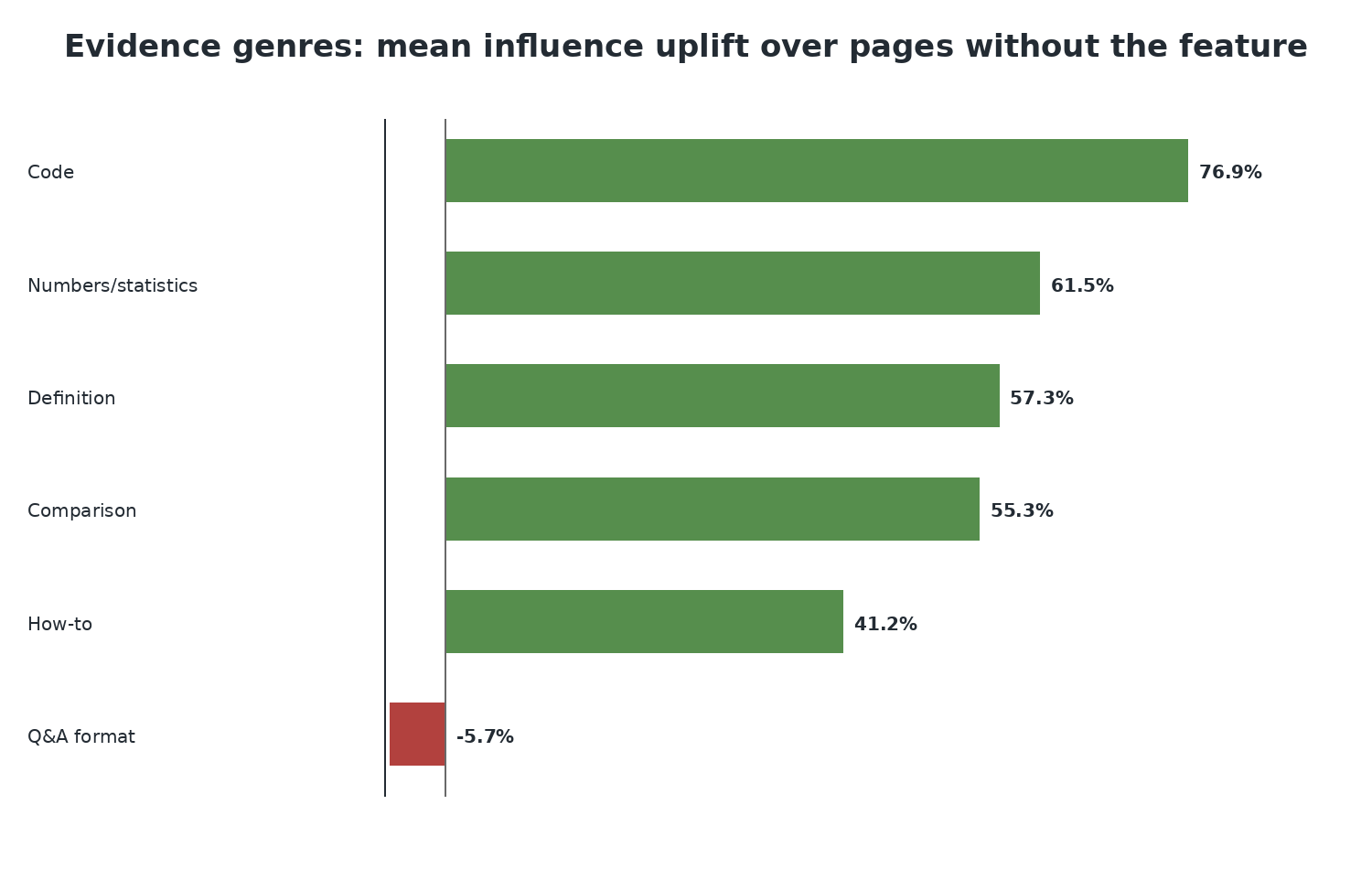}
\caption{Mean influence uplift associated with evidence genres. Q\&A format is the important negative case. Source: geo-citation-lab public report.}
\end{figure}

\begin{longtable}[]{@{}lrrr@{}}
\toprule\noalign{}
Feature & True mean & False mean & Relative difference \\
\midrule\noalign{}
\endhead
\bottomrule\noalign{}
\endlastfoot
Contains code & 0.1747 & 0.0988 & +76.88\% \\
Contains numbers/statistics & 0.1171 & 0.0725 & +61.55\% \\
Contains definition markers & 0.1252 & 0.0795 & +57.33\% \\
Contains comparison content & 0.1389 & 0.0894 & +55.28\% \\
Contains how-to content & 0.1296 & 0.0918 & +41.20\% \\
Q\&A format & 0.0947 & 0.1005 & -5.74\% \\
\end{longtable}

The evidence-genre table provides one of the paper's most actionable findings. Pages containing code, numbers, definitions, comparisons, or how-to content show higher mean influence. Q\&A format shows the opposite direction. The likely mechanism is that evidence genres create reusable support units, while Q\&A formatting is only a surface wrapper.

\subsection{Semantic Role and Usage Style}\label{semantic-role-and-usage-style}

\begin{figure}
\centering
\includegraphics[width=\linewidth]{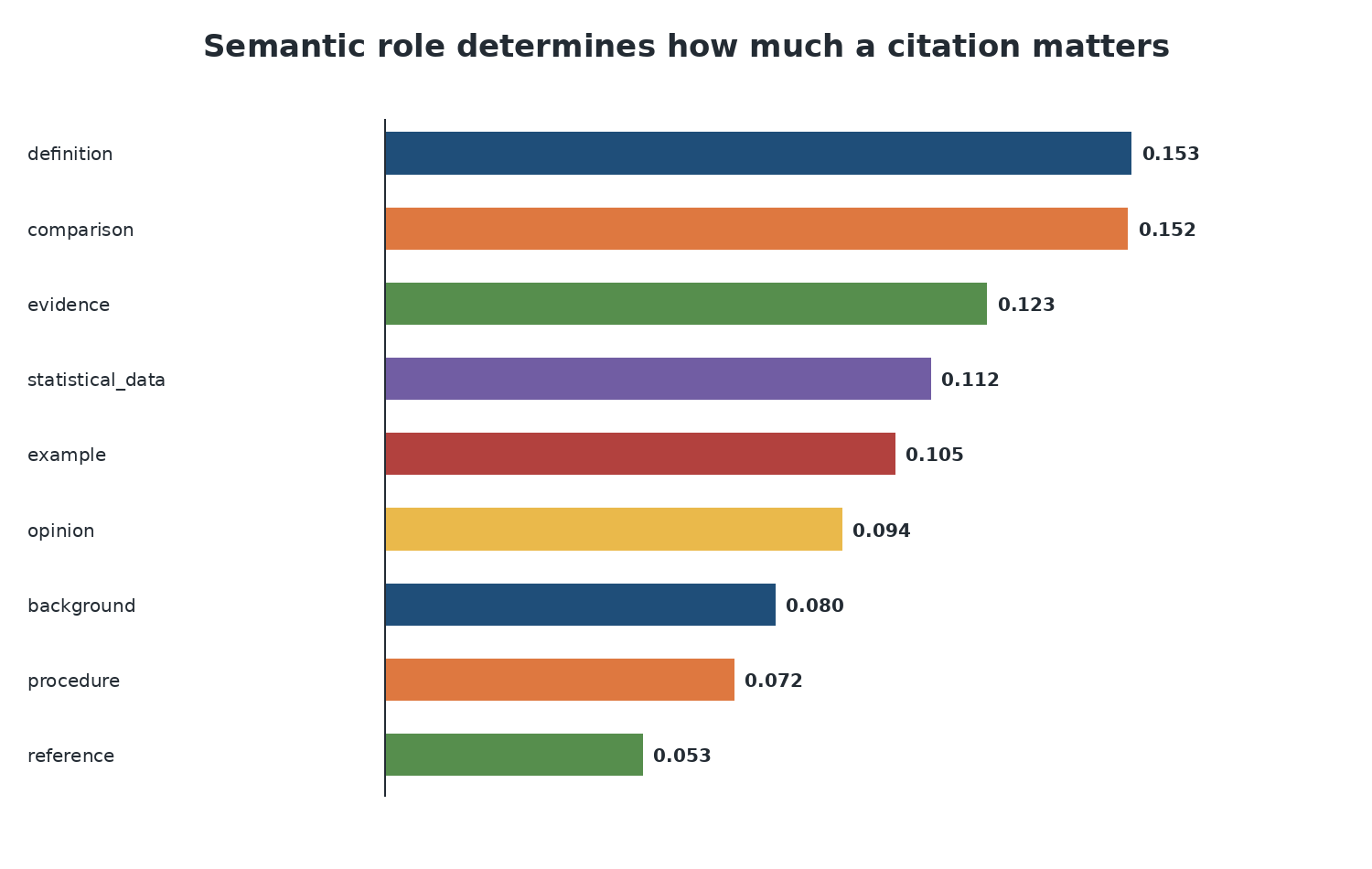}
\caption{Semantic role determines how much a citation matters. Source: geo-citation-lab public report.}
\end{figure}

\begin{longtable}[]{@{}lrr@{}}
\toprule\noalign{}
Semantic role & N & Mean influence \\
\midrule\noalign{}
\endhead
\bottomrule\noalign{}
\endlastfoot
definition & 1,663 & 0.1531 \\
comparison & 1,719 & 0.1524 \\
evidence & 6,216 & 0.1235 \\
statistical\_data & 504 & 0.1120 \\
example & 1,468 & 0.1047 \\
opinion & 846 & 0.0938 \\
background & 5,582 & 0.0801 \\
procedure & 497 & 0.0717 \\
reference & 1,298 & 0.0529 \\
\end{longtable}

Semantic role analysis reveals that the highest influence roles are definition and comparison. Reference-only citations are much weaker. This supports the absorption concept: a source gains answer-level importance when it supplies the form of information that the answer needs.

\begin{longtable}[]{@{}lrr@{}}
\toprule\noalign{}
Usage style & N & Mean influence \\
\midrule\noalign{}
\endhead
\bottomrule\noalign{}
\endlastfoot
fact\_source & 5,411 & 0.1241 \\
synthesized & 3,967 & 0.0964 \\
paraphrased & 5,305 & 0.0955 \\
background\_only & 5,100 & 0.0775 \\
\end{longtable}

The usage-style table suggests that direct factual support carries higher influence than background-only use. This is consistent with source attribution research: citation presence alone is too weak as an evaluation target.

\subsection{Domain Type and Platform-Specific Correlations}\label{domain-type-and-platform-specific-correlations}

\begin{longtable}[]{@{}lrr@{}}
\toprule\noalign{}
Domain type & N & Mean influence \\
\midrule\noalign{}
\endhead
\bottomrule\noalign{}
\endlastfoot
encyclopedia & 527 & 0.2144 \\
academic\_publishing & 86 & 0.1118 \\
commercial & 11,779 & 0.1028 \\
nonprofit & 2,009 & 0.0971 \\
academic & 1,024 & 0.0815 \\
government & 892 & 0.0769 \\
news\_media & 1,546 & 0.0726 \\
\end{longtable}

This table creates another selection-absorption split. News appears frequently in the candidate pool, yet \texttt{news\_media} has lower mean influence than encyclopedia and several other domain types in the absorption table. This suggests that answer engines often need explanatory pages after they have identified timely sources.

\begin{figure}
\centering
\includegraphics[width=\linewidth]{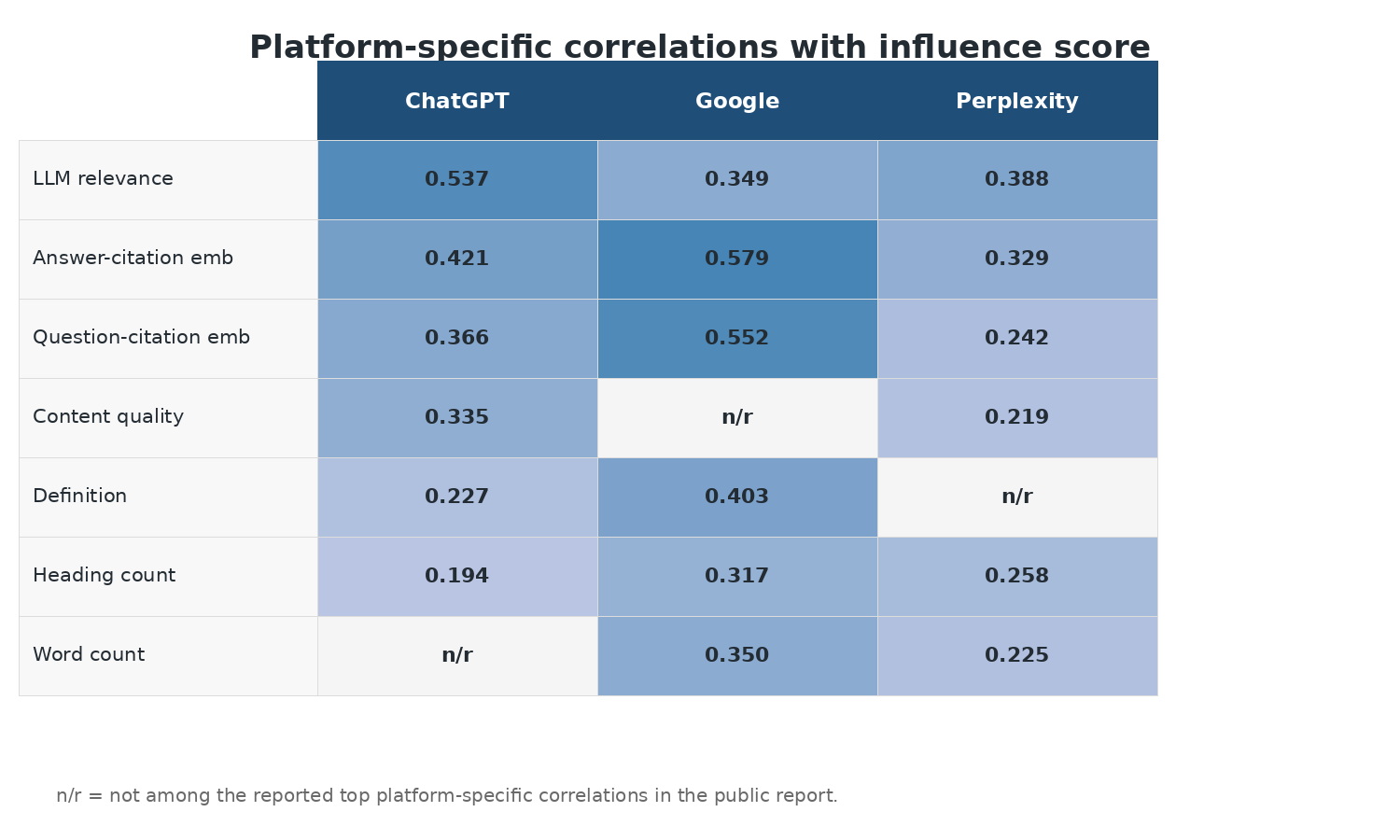}
\caption{Platform-specific correlations indicate that each engine weights semantic and structural features differently. Source: geo-citation-lab public report.}
\end{figure}

ChatGPT's strongest reported signal is LLM relevance. Google's strongest reported signals are answer-citation and question-citation embedding similarities, along with definition markers. Perplexity combines relevance with heading count and length. These patterns should be treated as platform-specific descriptive profiles, not permanent platform laws.

\subsection{Platform Strategy Profiles as Descriptive Archetypes}\label{platform-strategy-profiles-as-descriptive-archetypes}

The three platforms can be summarized as descriptive archetypes, while preserving the warning that product behavior may change. ChatGPT appears citation-sparse and absorption-heavy in this snapshot. This means that the platform tends to cite fewer sources per prompt, but the selected sources receive higher average influence under the constructed proxy. This pattern may reflect answer synthesis that uses a smaller evidence base more intensively. It may also reflect differences in citation rendering, answer length, browsing implementation, or the way the answer HTML is parsed by the pipeline.

Google AI Overview/Gemini appears broader in citation selection and lower in average citation absorption. It also shows strong response to explicit source-request prompts in the B layer and a pronounced English advantage in the C layer. This may reflect a search-product heritage where broad source exposure and snippet-like support are important. It may also reflect a citation interface that attaches more sources to a generated response, reducing average influence per source.

Perplexity appears citation-rich and coverage-oriented. It reaches the highest mean citations per prompt and triggers search for every observed prompt in the cleaned report. Its average absorption is closer to Google than to ChatGPT. This suggests a retrieval style that favors breadth, source diversity, and visible provenance. For GEO measurement, Perplexity-like behavior is useful because it exposes many candidate sources; for answer influence, it may require additional metrics to distinguish central sources from peripheral ones.

These archetypes should not be converted into permanent claims about vendor strategies. They are empirical profiles of the available snapshot. A longitudinal replication should report model version, data-collection window, prompt execution order, and UI mode where possible. If those details cannot be reconstructed, the limitation should remain visible in the paper rather than hidden in a footnote.

\subsection{Why the Negative Q\&A Result Matters}\label{why-the-negative-qa-result-matters}

The Q\&A result is especially important because many practical GEO guides recommend FAQ-style content as a general solution. In the reported means, Q\&A pages have slightly lower influence than non-Q\&A pages. This does not prove that FAQ content is harmful. It shows that the surface format alone is not a sufficient signal of answer usefulness.

A plausible explanation is that Q\&A formatting often creates short, isolated responses. Such pages may answer narrow questions but fail to provide the evidence density needed for synthesis across a more complex user prompt. Another explanation is taxonomy noise: some FAQ pages are thin support pages, and some non-Q\&A pages are detailed explainers, government documents, or encyclopedic resources. A follow-up regression should control for word count, structure, quality, relevance, and domain type before drawing a sharper conclusion.

The conservative scientific conclusion is therefore narrow and useful: do not treat FAQ conversion as a universal GEO intervention. If Q\&A content is used, it should contain substantive definitions, quantitative evidence, comparison logic, and clear section structure. The value comes from the evidence inside the page, not from the presence of question marks in headings.

\subsection{Industry and Question Type}\label{industry-and-question-type}

\begin{figure}
\centering
\includegraphics[width=\linewidth]{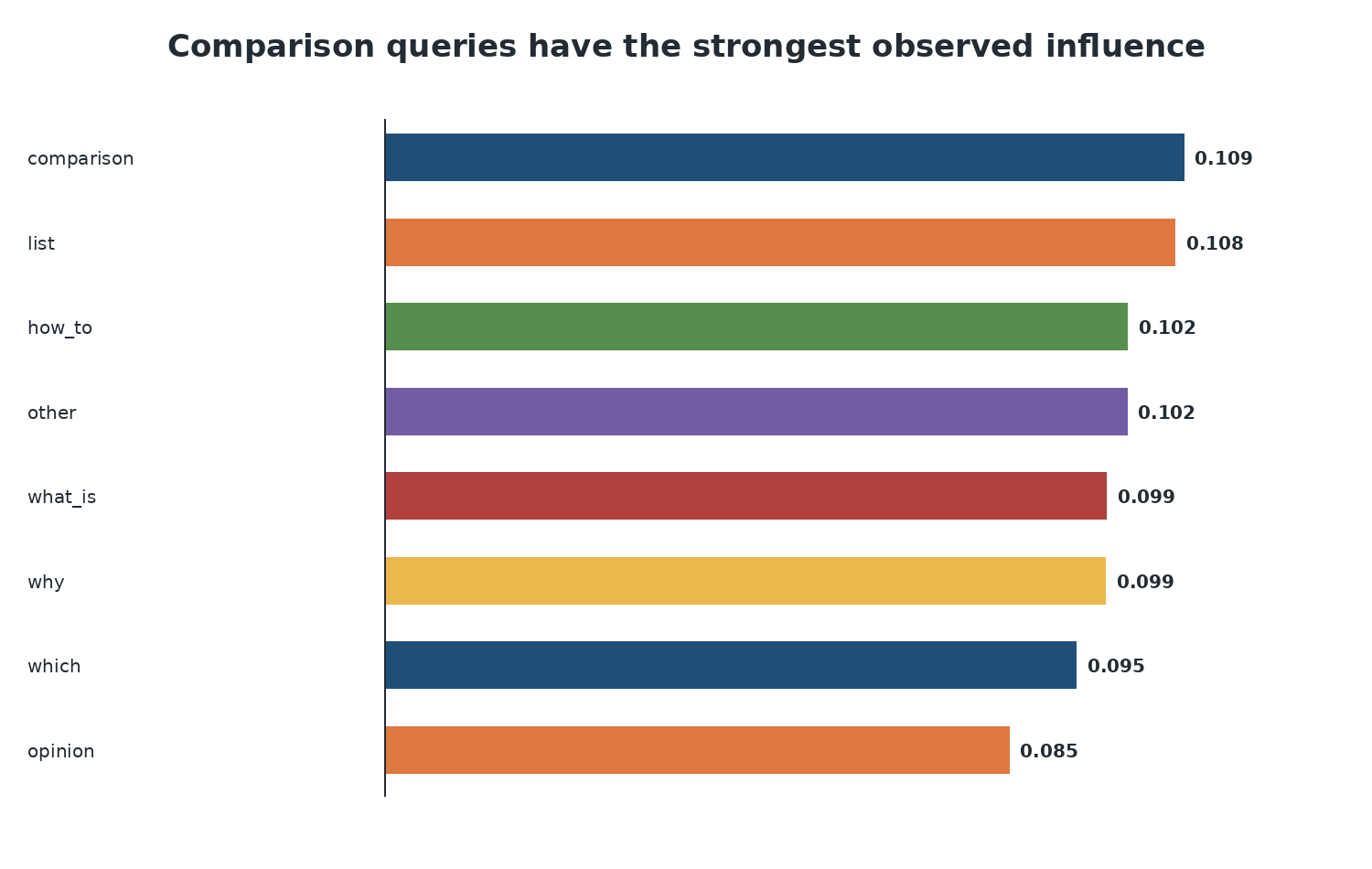}
\caption{Mean influence by question type. Source: geo-citation-lab public report.}
\end{figure}

\begin{longtable}[]{@{}lrr@{}}
\toprule\noalign{}
Industry & N & Mean influence \\
\midrule\noalign{}
\endhead
\bottomrule\noalign{}
\endlastfoot
A\_technology & 2,252 & 0.1272 \\
A\_healthcare & 2,379 & 0.1021 \\
A\_commerce & 2,243 & 0.0994 \\
A\_finance & 1,831 & 0.0965 \\
A\_news & 2,204 & 0.0948 \\
A\_local & 2,223 & 0.0916 \\
\end{longtable}

Technology has the highest reported average influence among industries, followed by healthcare and commerce. Comparison questions have the highest reported average influence among question types. These categories naturally demand definitions, criteria, contrastive evidence, and structured explanation, which align with the evidence-container model.

\section{Mechanistic Interpretation: The Evidence-Container Hypothesis}\label{mechanistic-interpretation-the-evidence-container-hypothesis}

The evidence-container hypothesis states that a page becomes valuable to a generative engine when it can be decomposed into reusable, semantically aligned information units. A high-quality evidence container has clear topical scope, section headings that mirror likely user subquestions, paragraphs that each carry a distinct claim, and extractable evidence such as definitions, statistics, comparisons, examples, and step sequences.

This hypothesis explains why word count alone is an incomplete proxy. A long page full of boilerplate, navigation, repetition, or unrelated material may not be useful. A long page with modular structure gives the engine multiple chances to map user intent to specific answer segments. The evidence-container frame also explains why Q\&A formatting does not automatically help. A Q\&A page can be shallow, sparse, redundant, or poorly aligned with the generated answer.

The selection layer still matters. A page that is never retrieved or never selected cannot be absorbed. Authority, language, region, source type, and platform-specific preferences shape the gateway. Once inside the gateway, absorption depends more heavily on semantic alignment and usable evidence structure.

\subsection{Evidence-Container Design: Operational Criteria}\label{evidence-container-design-operational-criteria}

The evidence-container hypothesis can be translated into operational criteria for future measurement. A page should be scored on topical scope, section modularity, evidence density, source transparency, and semantic alignment. Topical scope asks whether the page has a clearly bounded subject rather than a diffuse collection of loosely related claims. Section modularity asks whether headings divide the page into reusable answer units. Evidence density asks whether the page contains definitions, statistics, examples, comparisons, procedural steps, caveats, and source links. Source transparency asks whether facts are traceable to primary or reputable references. Semantic alignment asks whether the page maps to the likely user tasks implied by the prompt set.

These criteria can be measured manually in a small validation sample and automatically in the full dataset. Manual annotation is valuable because automatic features may miss context. For example, a page can contain many numbers without providing useful statistics, or it can contain many headings that repeat generic marketing claims. A high-quality validation protocol would sample pages from each influence quartile, each platform, and each domain type. Annotators would label whether each page contains answer-ready definitions, decision criteria, comparative evidence, and numerical support.

A second validation layer should examine answer sentences. For each generated answer, sentence-level support can be linked to cited pages. This would distinguish a page that merely appears in the citation list from a page that supports multiple answer claims. The current influence score approximates this distinction through paragraph coverage and text similarity. Sentence-level annotation would provide an independent benchmark for evaluating and improving the influence proxy.

\subsection{Practical GEO Metrics Derived From the Framework}\label{practical-geo-metrics-derived-from-the-framework}

A publisher-facing GEO dashboard should avoid a single visibility number. A more robust dashboard would include at least five metrics. Selection rate measures whether a page or domain appears in citation pools for target prompts. Citation breadth measures how many citations appear per prompt and how frequently the domain recurs. Absorption score measures answer-level influence. Support quality measures whether cited pages actually substantiate generated claims. Coverage equity measures whether visibility is concentrated among a small set of domains or distributed across credible long-tail sources.

The same dashboard should separate prompt families. A page that performs well for ``what is'' prompts may not perform well for comparison, how-to, or high-risk prompts. The question-type results show that comparison prompts have relatively high mean influence. This suggests that comparison-oriented content should be evaluated under comparison prompts rather than under generic brand or definition prompts.

Finally, the dashboard should include time. Generative engines evolve quickly. A one-time GEO audit is informative but incomplete. Longitudinal measurement should repeat prompt panels at fixed intervals, log model and UI context, and track whether citation pools, influence scores, and source-type distributions drift over time.

\subsection{Practical Implication Under Scientific Constraints}\label{practical-implication-under-scientific-constraints}

For selection, build credibility, make pages fetchable, align metadata and titles with likely query intent, and publish in environments that engines already treat as reliable. For absorption, write pages as modular evidence containers with definitions, numbers, comparisons, and procedures. Use headings to expose the semantic skeleton of the page. For evaluation, track both citation occurrence and answer-level influence. A dashboard that counts citations alone will miss the central pattern of this dataset.

\subsection{Why This Differs From Classical SEO Advice}\label{why-this-differs-from-classical-seo-advice}

Classical SEO is still relevant because source authority and domain recognizability influence the gateway into citation pools. The new layer is answer participation. GEO content must be easy to index, easy to retrieve, easy to quote or paraphrase, and easy to map into the user's actual task. These requirements make content design closer to information architecture and evidence packaging than to keyword density.

The data also discourages a single universal optimization recipe. ChatGPT, Google, and Perplexity behave differently. The same prompt style or language condition can increase citation breadth on one engine and fail to do so on another. A high-level GEO theory therefore needs platform interactions, not only global ranking factors.

\section{Threats to Validity}\label{threats-to-validity}

\subsection{Internal Validity}\label{internal-validity}

Construct validity of \texttt{influence\_score} is the central threat. The score is a plausible absorption proxy, but it is not direct model attention or a causal trace. It combines observable text and citation-position features into a single number. This gives the study a measurable outcome, yet it also means that outcome interpretation must remain tied to the formula.

Outcome-component leakage is another risk. Any model that predicts \texttt{influence\_score} using its own components would be circular. This manuscript explicitly avoids that error by distinguishing outcome components from permissible explanatory variables.

Fetch-ok selection is also important. Absorption analysis excludes failed fetches. If failed pages systematically differ from successful pages, absorption estimates are conditional on fetchability. A follow-up analysis should model \texttt{fetch\_ok} as an intermediate selection outcome.

The lack of unified record-level timestamps limits temporal interpretation. AI search products change quickly. Without aligned timestamps, the dataset should be interpreted as a static research snapshot rather than a live platform-monitoring source.

\subsection{External Validity}\label{external-validity}

The 602 prompts are designed, not randomly sampled from all user behavior. The language contrast focuses on Chinese and English; results should not be generalized to all non-English contexts. The source-type taxonomy contains unknown and noisy values. Country and language shares are reported only after excluding unknown values. AI search products may change retrieval policies, citation rendering, browsing integrations, and model backends after the data snapshot.

\subsection{Statistical Validity}\label{statistical-validity}

The current manuscript uses descriptive statistics from the public report. A confirmatory microdata analysis should compute uncertainty intervals, correct for multiple comparisons, cluster errors by prompt and domain, and conduct robustness checks under alternative influence-score weights. The absence of these inferential quantities is a deliberate scientific choice in the current manuscript, not an omission hidden as certainty.

\section{Robustness and Validation Plan}\label{robustness-and-validation-plan}

\begin{longtable}[]{@{}
  >{\raggedright\arraybackslash}p{(\columnwidth - 4\tabcolsep) * \real{0.3333}}
  >{\raggedright\arraybackslash}p{(\columnwidth - 4\tabcolsep) * \real{0.3333}}
  >{\raggedright\arraybackslash}p{(\columnwidth - 4\tabcolsep) * \real{0.3333}}@{}}
\toprule\noalign{}
\begin{minipage}[b]{\linewidth}\raggedright
Check
\end{minipage} & \begin{minipage}[b]{\linewidth}\raggedright
Purpose
\end{minipage} & \begin{minipage}[b]{\linewidth}\raggedright
Expected output
\end{minipage} \\
\midrule\noalign{}
\endhead
\bottomrule\noalign{}
\endlastfoot
Alternative influence weights & Test whether main absorption findings depend on the 0.20/0.15/0.20/0.25/0.20 weighting scheme. & Rank stability, platform means, quartile comparisons. \\
Content-only influence proxy & Remove \texttt{ref\_count} and \texttt{first\_position} components to test whether results survive without citation-display features. & Correlation and platform comparison tables. \\
Domain-level deduplication & Prevent high-frequency domains from dominating citation-level records. & Prompt-domain and domain-level summaries. \\
Remove mega-domains & Test sensitivity to Wikipedia, YouTube, Reddit, and major news/platform domains. & Recomputed selection and absorption summaries. \\
Fetch-failure model & Predict \texttt{fetch\_ok} from source type, platform, and domain to diagnose selection into the absorption sample. & Logistic model and missingness report. \\
Two-way clustered inference & Account for repeated prompts and repeated domains. & Robust standard errors and confidence intervals. \\
\end{longtable}

A fully reproducible research release should attach a notebook or script that starts from the repository CSV files and exports every table and figure. The manuscript should also include a provenance table indicating which claims are direct observations, which are descriptive aggregations, and which are interpretive hypotheses.

\subsection{Confirmatory Analysis Plan}\label{confirmatory-analysis-plan}

The confirmatory analysis should be specified at the level of code and model design before inferential claims are reported. The primary selection outcome should be prompt-platform citation count. The primary absorption outcome should be influence score restricted to fetch-ok pages. Secondary outcomes should include selection trigger, domain-type category, semantic role, usage style, and content-genre indicators.

The first confirmatory table should estimate platform differences in citation count using a negative binomial model, because citation counts are non-negative integers and likely overdispersed. The second confirmatory table should estimate absorption differences using fractional logit or beta regression, because influence score is bounded between 0 and 1. The third table should test whether structural and semantic features remain associated with absorption after controlling for platform, domain type, industry, question type, and language. Standard errors should be clustered at minimum by prompt, and ideally by both prompt and domain.

The primary robustness family should alter the outcome definition. One version should remove citation-display components from the influence score and retain only content-overlap or semantic-coverage components. Another version should use a top-quartile binary absorption outcome. A third version should aggregate to prompt-domain level so that repeated citations from the same prompt do not dominate. A fourth version should remove mega-domains such as Wikipedia, YouTube, Reddit, and large news portals.

The pre-registration should also define exclusion rules. Failed fetches should be retained for selection analysis and excluded only for absorption analysis. Unknown country and language values should not be silently dropped; tables should state the identifiable denominator. No variable that directly appears inside the influence-score formula should be used as an independent explanatory variable in the primary absorption model.

\section{Responsible GEO and Ethics}\label{responsible-geo-and-ethics}

GEO can be used to improve the discoverability of accurate, well-structured information. It can also be misused to manipulate generated answers, flood engines with low-quality synthetic content, or exploit citation mechanisms. This paper recommends responsible GEO: make content more accurate, transparent, fetchable, and useful for verification rather than designing hidden instructions or deceptive artifacts.

The evidence-container strategy is ethically acceptable when it improves clarity, citation traceability, and user understanding. It becomes harmful when it fabricates authority, injects misleading numbers, hides prompt-injection text, or attempts to steer models away from competing evidence. Future GEO benchmarks should include abuse-resistance and citation-quality dimensions, not only visibility.

Because generative engines mediate public access to information, source-selection concentration also raises fairness questions. High authority and US-English dominance may disadvantage smaller, local, multilingual, or long-tail sources. Measurement frameworks should therefore evaluate both accuracy and equitable visibility.

\section{Reproducibility and Data Availability}

This study is based on the public geo-citation-lab repository and its associated public report. The repository provides the dataset and analysis pipeline used to study how AI search engines select and use citations. The present manuscript reports descriptive quantities that are traceable to that public source and distinguishes them from confirmatory models that require a fresh raw-CSV rerun.

A fully reproducible research release should include a version-pinned analysis script that starts from the repository CSV files, regenerates all descriptive tables and figures, records model versions for embedding and LLM-scoring steps, and exports the confirmatory inference tables described in the methodology section. Failed fetches should remain in the selection analysis and be excluded only for absorption analysis, with denominators reported explicitly.

The project repository is \url{https://github.com/yaojingang/geo-citation-lab}. Repository and public-report links are retained for source traceability.

\section{Conclusion}\label{conclusion}

This paper argues that GEO should be measured as a two-stage process: source selection followed by source absorption. The geo-citation-lab dataset supports this distinction. Search triggering is near universal across the studied platforms, yet citation breadth differs substantially. More importantly, breadth and absorption diverge: ChatGPT cites fewer sources but has much higher mean influence among fetched citations, while Perplexity and Google cite broader source sets with lower mean per-source absorption.

The strongest descriptive evidence points toward semantic alignment and evidence structure. High-influence pages are longer, more modular, and more semantically aligned with generated answers. Pages containing definitions, numerical information, comparison content, code, and procedural content show higher mean influence. Q\&A format alone does not help in the reported means.

The paper's main scientific contribution is a measurement scaffold that separates exposure from answer influence. The main practical implication is evidence-container design. The main methodological warning is that descriptive GEO findings should not be prematurely converted into universal causal rules. Follow-up work should rerun the raw CSV pipeline, add uncertainty estimates, and test whether controlled content rewrites can causally improve both selection and absorption.

\appendix
\section{Claim-Level Self-Audit}\label{appendix-a.-claim-level-self-audit}

\begin{longtable}[]{@{}
  >{\raggedright\arraybackslash}p{(\columnwidth - 6\tabcolsep) * \real{0.2500}}
  >{\raggedright\arraybackslash}p{(\columnwidth - 6\tabcolsep) * \real{0.2500}}
  >{\raggedright\arraybackslash}p{(\columnwidth - 6\tabcolsep) * \real{0.2500}}
  >{\raggedright\arraybackslash}p{(\columnwidth - 6\tabcolsep) * \real{0.2500}}@{}}
\toprule\noalign{}
\begin{minipage}[b]{\linewidth}\raggedright
Claim
\end{minipage} & \begin{minipage}[b]{\linewidth}\raggedright
Evidence status
\end{minipage} & \begin{minipage}[b]{\linewidth}\raggedright
Limitation
\end{minipage} & \begin{minipage}[b]{\linewidth}\raggedright
Recommended follow-up action
\end{minipage} \\
\midrule\noalign{}
\endhead
\bottomrule\noalign{}
\endlastfoot
Three platforms nearly always trigger search. & Direct public report table. & Near-ceiling rates limit logistic-model variation. & Report exact denominators and treat as descriptive. \\
Citation breadth differs by platform. & Direct public report table. & Citation count is sensitive to platform UI and deduplication rules. & Add prompt-domain deduplication robustness. \\
ChatGPT has higher mean absorption among fetched pages. & Direct public report absorption summary. & \texttt{influence\_score} is a constructed proxy and conditional on fetch\_ok. & Run alternative influence definitions and fetch-ok selection model. \\
High-influence pages are longer and more structured. & Direct top/bottom quartile comparison. & Length and structure may be correlated with domain type and industry. & Add multivariate fractional model with controls. \\
Semantic relevance is the strongest independent reported correlation. & Direct correlation table. & LLM scoring model and embedding model details should be versioned. & Document scoring prompts/models and rerun if possible. \\
Q\&A format does not improve absorption. & Direct boolean-feature mean comparison. & Q\&A definition may be broad; possible confounding with low-quality FAQ pages. & Inspect examples and add controls for quality and relevance. \\
News is frequent but not deepest absorbed. & Combination of source-type frequency and domain-type influence table. & Domain taxonomy may contain noise and platform composition effects. & Rerun with standardized taxonomy and platform fixed effects. \\
Evidence-container hypothesis explains patterns. & Interpretive synthesis from multiple descriptive results. & Hypothesis is not causally proven. & Test with controlled page rewrites and randomized intervention study. \\
\end{longtable}

\section{Proposed Experimental Extension}\label{appendix-b.-proposed-experimental-extension}

A natural research extension is a controlled intervention study. Select a stratified sample of pages across industries and generate matched rewrites that vary only one feature family at a time: structure, evidence density, semantic alignment, source transparency, and page length. Submit matched prompts to the same platforms at the same timestamp window and compare selection and absorption outcomes.

A minimal randomized design would use five arms: original page, structured headings only, evidence blocks only, semantic-intent alignment only, and full evidence-container rewrite. Outcomes would include citation probability, influence score, citation role, sentence-level support quality, and user-visible citation placement. This design would convert the present descriptive paper into a causal GEO study.

\section{Compact Data Dictionary for the Reproduction Script}\label{appendix-c.-compact-data-dictionary-for-the-reproduction-script}

\begin{longtable}[]{@{}
  >{\raggedright\arraybackslash}p{(\columnwidth - 6\tabcolsep) * \real{0.2500}}
  >{\raggedright\arraybackslash}p{(\columnwidth - 6\tabcolsep) * \real{0.2500}}
  >{\raggedright\arraybackslash}p{(\columnwidth - 6\tabcolsep) * \real{0.2500}}
  >{\raggedright\arraybackslash}p{(\columnwidth - 6\tabcolsep) * \real{0.2500}}@{}}
\toprule\noalign{}
\begin{minipage}[b]{\linewidth}\raggedright
Field family
\end{minipage} & \begin{minipage}[b]{\linewidth}\raggedright
Example fields
\end{minipage} & \begin{minipage}[b]{\linewidth}\raggedright
Role in analysis
\end{minipage} & \begin{minipage}[b]{\linewidth}\raggedright
Caution
\end{minipage} \\
\midrule\noalign{}
\endhead
\bottomrule\noalign{}
\endlastfoot
Prompt metadata & layer, industry, question type, language, style & Independent design variables. & Prompt taxonomy must be normalized across platform files. \\
Platform metadata & platform, prompt id, response id & Stratification and fixed effects. & Platform behavior is time-sensitive. \\
Search-layer citation & citation domain, citation URL, search triggered & Selection outcomes and source pool. & Deduplication should be tested at URL, domain, and prompt-domain levels. \\
Source metadata & website type, country, language, Final\_DR & Source-selection covariates. & Unknown values require explicit denominator reporting. \\
Fetch status & fetch\_ok, fetched\_html, error state & Absorption-sample inclusion and missingness. & Failed fetches may be non-random. \\
Page structure & word count, headings, paragraphs, list density & Candidate absorption predictors. & Structure may proxy for editorial quality. \\
Evidence genre & definitions, numbers, comparisons, how-to, code, Q\&A & Evidence-container features. & Binary detectors need validation on examples. \\
Semantic alignment & embedding similarity, LLM relevance, LLM quality & Core absorption predictors. & Model and prompt versions should be logged. \\
Influence components & ref\_count, first\_position\_ratio, coverage, TF-IDF, n-gram overlap & Outcome construction. & Avoid using these as predictors of the same outcome. \\
\end{longtable}

\section{Reproducibility and Release Checklist}\label{appendix-d.-reproducibility-and-release-checklist}

This checklist summarizes the manuscript's reproducibility posture. It is written as a release-oriented research checklist rather than as a platform-specific filing checklist.

\begin{longtable}[]{@{}
  >{\raggedright\arraybackslash}p{(\columnwidth - 6\tabcolsep) * \real{0.3000}}
  >{\raggedright\arraybackslash}p{(\columnwidth - 6\tabcolsep) * \real{0.4500}}
  >{\raggedright\arraybackslash}p{(\columnwidth - 6\tabcolsep) * \real{0.2500}}@{}}
\toprule\noalign{}
\begin{minipage}[b]{\linewidth}\raggedright
Item
\end{minipage} & \begin{minipage}[b]{\linewidth}\raggedright
Current handling
\end{minipage} & \begin{minipage}[b]{\linewidth}\raggedright
Status
\end{minipage} \\
\midrule\noalign{}
\endhead
\bottomrule\noalign{}
\endlastfoot
Author metadata & Author order is Zhang Kai first, He Xinyue second, and Yao Jingang third; Yao public profile links and repository links are retained for traceability. & Included \\
Public dataset attribution & The geo-citation-lab repository and public report are cited as the data source. & Included \\
Evidence boundaries & Descriptive claims, interpretive hypotheses, and proposed confirmatory models are separated in the text. & Included \\
Denominator discipline & Search-layer and fetch-ok denominators are reported separately. & Included \\
Circularity control & Influence-score components are excluded from primary explanatory language. & Included \\
Microdata inference & Bootstrap intervals, clustered standard errors, and multivariate models are specified as a follow-up microdata analysis. & Recommended extension \\
Temporal replication & Repeating the prompt battery over time is specified as a longitudinal robustness extension. & Recommended extension \\
\end{longtable}

\section{Additional Claim-Level Notes}\label{appendix-e.-additional-claim-level-notes}

The paper's central empirical contrast can be independently checked from two tables in the public report: the search-layer platform summary and the fetch-ok absorption summary. The search-layer table establishes citation breadth. The absorption table establishes mean and median influence among successfully fetched citations. The contrast between these two tables motivates the selection-absorption framework.

The evidence-container hypothesis can be checked from three additional descriptive blocks: the top-versus-bottom quartile profile, the independent feature correlations, and the evidence-genre mean differences. The hypothesis gains credibility because the same direction appears across length, headings, paragraphs, list density, semantic similarity, LLM relevance, definitions, numbers, comparisons, and how-to indicators. The Q\&A exception prevents over-simplified format advice and strengthens the argument that evidence density is more important than surface layout.

The strongest remaining weakness is causal identification. A reviewer could reasonably ask whether high-influence pages are selected because they are high-quality, because they belong to favored domains, because they are longer, because they have better semantic alignment, or because the platform's retrieval component surfaced them earlier. Follow-up work should answer this with multivariate controls and controlled rewrites. Until then, the manuscript should be read as a rigorous measurement paper and hypothesis generator.

\section{References}\label{references}

{[}1{]} Aggarwal, P., Murahari, V., Rajpurohit, T., Kalyan, A., Narasimhan, K., and Deshpande, A. (2024). \emph{GEO: Generative Engine Optimization}. arXiv:2311.09735. Accepted to KDD 2024.

{[}2{]} Chen, M., Wang, X., Chen, K., and Koudas, N. (2025). \emph{Generative Engine Optimization: How to Dominate AI Search}. arXiv:2509.08919.

{[}3{]} Tian, Z., Chen, Y., Tang, Y., Liu, J., and Jia, R. (2026). \emph{Diagnosing and Repairing Citation Failures in Generative Engine Optimization}. arXiv:2603.09296.

{[}4{]} Liu, Z., and Xu, P. (2026). \emph{Think Before Writing: Feature-Level Multi-Objective Optimization for Generative Citation Visibility}. arXiv:2604.19113.

{[}5{]} Yuan, J., Wang, J., Wang, Z., Sun, Q., Wang, R., and Li, J. (2026). \emph{AgenticGEO: A Self-Evolving Agentic System for Generative Engine Optimization}. arXiv:2603.20213.

{[}6{]} Yu, J., Yang, M., Ding, Y., and Sato, H. (2026). \emph{Structural Feature Engineering for Generative Engine Optimization: How Content Structure Shapes Citation Behavior}. arXiv:2603.29979.

{[}7{]} Narayanan Venkit, P., Laban, P., Zhou, Y., Mao, Y., and Wu, C.-S. (2024). \emph{Search Engines in an AI Era: The False Promise of Factual and Verifiable Source-Cited Responses}. arXiv:2410.22349.

{[}8{]} Yang, K.-C. (2025). \emph{News Source Citing Patterns in AI Search Systems}. arXiv:2507.05301.

{[}9{]} Xu, Y., Qi, P., Chen, J., Liu, K., Han, R., Liu, L., Min, B., Castelli, V., Gupta, A., and Wang, Z. (2025). \emph{CiteEval: Principle-Driven Citation Evaluation for Source Attribution}. arXiv:2506.01829.

{[}10{]} Qian, H., Fan, Y., Zhang, R., and Guo, J. (2024). \emph{On the Capacity of Citation Generation by Large Language Models}. arXiv:2410.11217.

{[}11{]} Kirsten, E., Grosse Perdekamp, J., Upadhyay, M., Gummadi, K. P., and Zafar, M. B. (2025). \emph{Characterizing Web Search in The Age of Generative AI}. arXiv:2510.11560.

{[}12{]} Lewis, P., Perez, E., Piktus, A., Petroni, F., Karpukhin, V., Goyal, N., et al. (2020). \emph{Retrieval-Augmented Generation for Knowledge-Intensive NLP Tasks}. Advances in Neural Information Processing Systems.

{[}13{]} Karpukhin, V., Oguz, B., Min, S., Lewis, P., Wu, L., Edunov, S., Chen, D., and Yih, W.-t. (2020). \emph{Dense Passage Retrieval for Open-Domain Question Answering}. EMNLP.

{[}14{]} Nakano, R., et al. (2021). \emph{WebGPT: Browser-assisted question-answering with human feedback}. arXiv:2112.09332.

{[}15{]} Menick, J., et al. (2022). \emph{Teaching language models to support answers with verified quotes}. arXiv:2203.11147.

{[}16{]} geo-citation-lab repository. (2026). \emph{A dataset and analysis pipeline for studying how AI search engines select and use citations}. GitHub: \url{https://github.com/yaojingang/geo-citation-lab}. Accessed April 28, 2026.

{[}17{]} geo-citation-lab final report. (2026). \emph{Overseas GEO Research Long Report, recalculated version}. \url{https://yaojingang.github.io/geo-citation-lab/04-repet/final_report.html}. Accessed April 28, 2026.

{[}18{]} Yao Jingang. (2026). \emph{GitHub profile}. \url{https://github.com/yaojingang}. Accessed April 29, 2026.

{[}19{]} Yao Jingang. (2026). \emph{X profile}. \url{https://x.com/yaojingang}. Accessed April 29, 2026.

\end{document}